\documentclass[a4paper,fleqn,usenatbib]{mnras}
\usepackage[T1]{fontenc}
\usepackage{ae,aecompl}
\usepackage{graphicx}	
\usepackage{amssymb}	
\usepackage{amsbsy}
\usepackage{float}
\usepackage{color}
\usepackage{booktabs,caption}
\usepackage{amsmath} 
\usepackage{ulem}

\usepackage{mathrsfs,bm}
\newcommand{\bcdot}{\ensuremath{%
  \mathchoice%
   {\mskip\thinmuskip\lower0.2ex\hbox{\scalebox{1.5}{$\cdot$}}\mskip\thinmuskip}}%
   {\mskip\thinmuskip\lower0.2ex\hbox{\scalebox{1.5}{$\cdot$}}\mskip\thinmuskip}%
   {\lower0.3ex\hbox{\scalebox{1.2}{$\cdot$}}}%
   {\lower0.3ex\hbox{\scalebox{1.2}{$\cdot$}}}%
}

\def\del#1{{}}

\newcommand{\bnabla}{\ensuremath{\boldsymbol{\nabla}}}

\newcommand{\Msun}{M$_\odot$}

\newcommand{\Zsun}{Z$_\odot$}
\newcommand{\frand}{$f_{\rm{rand}}$}

\newcommand{\rhomax}{$\rho_{\rm{max}}$}

\title[How CRs mediate ISM evolution]{How Cosmic Rays Mediate the Evolution of the Interstellar Medium}

\author[C.M.~Simpson et al.]  {Christine~M.~Simpson$^{1,2}$\thanks{E-mail: csimpson@astro.uchicago.edu}, R\"udiger~Pakmor$^3$, Christoph~Pfrommer$^4$,
\newauthor Simon~C.~O.~Glover$^5$ \& Rowan Smith$^6$\\
$^{1}$Enrico Fermi Institute, The University of Chicago, Chicago, IL 60637, USA\\
$^{2}$Department of Astronomy \& Astrophysics, The University of Chicago, Chicago, IL 60637, USA\\
$^{3}$Max-Planck-Institut f\"{u}r Astrophysik, Karl-Schwarzschild-Str. 1, 85748 Garching, Germany\\
$^{4}$Leibniz-Institute for Astrophysics Potsdam (AIP), An der Sternwarte 16, 14482 Potsdam, Germany\\
$^5$ Universit\"{a}t Heidelberg, Zentrum f\"{u}r Astronomie, Institut f\"{u}r Theoretische Astrophysik,\\ Albert-Ueberle-Stra{\ss}e 2, 69120 Heidelberg, Germany\\
$^6$ Jodrell Bank Centre for Astrophysics, Department of Physics and Astronomy, University of Manchester, \\Oxford Road, Manchester M13 9PL, UK}
 
\pubyear{2022}

\begin{document}

\label{firstpage}
\pagerange{\pageref{firstpage}--\pageref{lastpage}}

\maketitle

\begin{abstract}

We explore the impact of diffusive cosmic rays (CRs) on the evolution of the interstellar medium (ISM) under varying assumptions of supernova explosion environment. In practice, we systematically vary the relative fractions of supernovae (SN) occurring in star-forming high-density gas and those occurring in random locations decoupled from star-forming gas to account for SN from run-away stars or explosions in regions that have been cleared by prior SN, stellar winds, or radiation.  We explore various mixed models by adjusting these fractions relative to each other. We find that in the simple system of a periodic stratified gas layer the ISM structure will evolve to one of two solutions: a ``peak driving'' state where warm gas is volume filling or a ``thermal runaway'' state where hot gas is volume filling.  CR pressure and transport are important factors that strongly influence the solution state the ISM reaches and have the ability to flip the ISM between solutions.  Observable signatures such as gamma ray emission and HI gas are explored. We find that gamma ray luminosity from pion decay is largely consistent with observations for a range of model parameters.  The thickness of the HI gas layer may be too compact, however, this may be due to a large cold neutral fraction of midplane gas.  The volume fraction of hot gas evolves to stable states in both solutions, but neither settles to a Milky Way-like configuration, suggesting that additional physics which is omitted here (e.g.\ a cosmological circum-galactic medium, radiation transport, or spectrally resolved and spatially varying CR transport) may be required.  

\end{abstract}
 

\section{Introduction}
Observations and modelling of the interstellar medium (ISM) have long indicated that the structure of the ISM is heavily influenced by the explosion of massive stars as supernovae (SN) \citep[e.g.][]{McKee1977}.  The phase structure of the ISM, where mass is aggregated in dense molecular clouds, embedded in a hot diffuse medium in pressure equilibrium, has been established through observations of neutral atomic and molecular gas \citep[e.g.][]{Cox2005} and observations of diffuse ionised gas (citation).

Modelling the phase structure and energetics of the ISM in a dynamic way has been a goal of simulations, both to understand the conditions for star formation within galaxies and to inform global models of galaxy evolution.  Simulations on the ISM scale have demonstrated the importance of SN explosion environment \citep[e.g.][]{Walch2015,Iffrig2015,Ohlin2019,Andersson2020}.  Simple one-dimensional models of SN shocks point to the importance of the background ISM gas density as a determining factor for how the SN shocks develop and the amount of energy and momentum they deposit in the ISM \citep{Draine}.  SN explosions that occur in dense molecular clouds will have most of their energy radiatively cooled on short timescales, but energy from explosions that occur in low density gas (due to pre-SN feedback, stellar kicks, or long SN delay times that allow for the dissipation of the natal molecular cloud) will remain in the ISM longer due to longer radiative cooling times \citep[e.g.][]{Walch2015b}.  

These simulations have most commonly studied the injection and dissipation of SN energy in a medium that only self-consistently tracks thermal and kinetic energies. However, the energy contained in the ISM comes in several other forms.  Indeed, the magnetic fields threading through the gas also contain energy, as do cosmic rays (CRs) in the medium, which are high energy protons, electrons, and other heavy atomic nuclei and represent an additional reservoir of energy.  Traditionally, these four energies are assumed to be in equilibrium \citep{BoularesCox1990,Cox2005,Naab2017}, but this assumption would have to be scrutinised with numerical simulations of the interstellar medium.

The non-thermal energies from magnetic fields and CRs are of particular interest because they transport and dissipate energy differently from the thermal gas.  CRs in particular have been posited as a mechanism for driving galactic outflows and as a regulator for star formation on interstellar scales \citep{Simpson2016,Girichidis2016a,Girichidis2018,Farber2018,Commercon2019} as well as on galactic scales in isolated galaxies \citep{Uhlig2012,Booth2013,Salem2014a,Pakmor2016b,Ruszkowski2017,Jacob2018,Butsky2018,Dashyan2020,Quataert2022,Girichidis2022,Farcy2022} and in cosmological settings \citep{Jubelgas2008,Salem2014b,Buck2020,Hopkins2021a}. The physics by which galaxies retain or expel gas and the efficiency with which they then turn gas into stars is of cosmological importance \citep[e.g.][]{Naab2017,Wechsler2018}, but the feedback processes that drive these behaviours are still uncertain.
SN explosions deposit only 5--15 per cent of their energy in CRs \citep{Caprioli2014,Pais2018}, but CR energy dissipates on longer timescales and couples to the thermal gas differently. 

Substantial simulation work has been done to explore how CRs on the ISM scale impact the local launching of outflows, but less has been done to explore how it affects the structure of the ISM.  Specifically, how CR pressure and transport impacts the picture of ISM structure driven by SN environment that has been built-up through the past decade.  \citet{Rathjen2021} have studied how a myriad of feedback effects including stellar winds, radiation, and CRs impact ISM structure and the development of outflows. They find that CRs only have a moderate impact on the multiphase structure of the ISM, but their models are only evolved for 100 Myr and they do not explore details of CR pressure or differences in transport.

The impact of ISM structure on observables associated with CRs, such as $\gamma$-ray emission or radio continuum emission, also have yet to be studied.
CR observables on galactic scales in the Milky Way \citep{Moskalenko1998,Evoli2008,Kissmann2015,Werhahn2021a,Hopkins2021b} and non-thermal radio and $\gamma$-ray emission from galaxies are a promising avenue for probing CR calorimetry \citep[which fraction of CR energy is radiated away as opposed to providing dynamical feedback,][]{Thompson2006,Lacki2010,Werhahn2021b,Werhahn2021c,Pfrommer2021} and of constraining the physics of CR transport \citep{Thomas2020}.

We aim to address these issues and specifically to explore how the presence of CRs impacts the structure and evolution of the ISM driven by SN explosions.  This paper is part of a series utilizing a stratified box setup to explore the impact of CRs on the ISM.  We test different models for SN explosion placement in the presence of CRs and demonstrate how the presence of an energy reservoir with different transport and dissipation timescales alters the ISM structure.  We quantify how CR energy is dissipated from the galactic midplane and explore models with different diffusivity properties.  We finally explore observable implications of our models in $\gamma$-ray and 21 cm emission.

The paper is organised as follows: Section \ref{sec:methods} describes our methodology and simulation set-up, including our model for SN explosion placement; Section \ref{sec:SNplacement} describes the effect SN explosion placement has on ISM structure when CRs are included; Sections \ref{sec:energy} and \ref{sec:energy_evo} describe the energetics of midplane gas; Section \ref{sec:kappa_diff} describes how our results depend on differences in the CR model; and Section \ref{sec:eff_pres} explores the impact of CRs on the midplane pressure.  In Section \ref{sec:obs} we look at observable signatures of these models in $\gamma$-ray and 21 cm emission.  Finally, we conclude with a discussion of the MW's ISM structure and how it compares to the models presented here.

\section{Methods}
\label{sec:methods}

We use the second-order accurate moving-mesh code {\small AREPO}  to study the impact of SN explosions and CR transport on the ISM \citep{Springel2010, Pakmor2016a}.   
{\small AREPO} gives a quasi-Lagrangian solution to the ideal (magneto-)hydrodynamics equations that combines advantages of an Eulerian scheme in capturing shocks and discontinuities and the advantages of a pure Lagrangian scheme with improved Galilean invariance.  
Here we describe the set-up and initial conditions of our simulations; a model for pressure in high density gas; additional gas physics modules such as self-gravity, radiative cooling, cosmic rays, and MHD; and the different schemes for SN explosion placement that we explore.

\subsection{Simulation Setup and Initial Conditions}

We simulate the impact of SN explosions on the ISM in a tall column of stratified gas meant to capture the behaviour of gas in a small patch of a galactic disc.  The simulation domain has a width of $L=1$ kpc in both the horizontal $x$ and $y$ directions and extends $\pm 5$ kpc in the vertical $z$ direction (for a box height of $L_z = 10$ kpc).  The domain has periodic boundaries in the $x$ and $y$ directions and outflow boundaries in the $z$ direction.    

Mass and energy can flow out of the simulation volume through the outflow boundaries at the top and bottom of the column.  
This is accomplished by fixing a layer of static `sticky' cells along these boundaries that do not move (i.e. their positions are not updated by the moving mesh algorithm).  
This is necessary to prevent the mesh structure (that normally moves with the motion of the thermal gas) from exiting the simulation volume.  
A layer of `ghost' cells are mirrored across the outflow boundaries and gas can move across this boundary in the outward direction only, leaving the simulation.

The initial positions and densities of gas cells are set according to an assumption of isothermal hydro-static equilibrium following \citet{Creasey2013} and used in \citet{Simpson2016}, the preliminary study to this work.  The initial density of gas cells depends on the choice of initial gas surface density, $\Sigma_0$, and gas fraction, $f_g$, as well as the cell's height $h$ above the midplane:
\begin{equation}
\label{eq:rho}
\rho_0(h) = \frac{\Sigma_0}{2b_0} \mathrm{sech}^2 \left( \frac{h}{b_0} \right),
\end{equation}
\begin{equation}
\label{eq:bscale}
b_0 = \frac{f_\rmn{g} k_\rmn{B} T_0}{m_\rmn{p} \mu_0 \pi G \Sigma_0}.
\end{equation}
\noindent The constant $G$ is the gravitational constant, $k_\rmn{B}$ is Boltzmann's constant, and $m_\rmn{p}$ is the proton mass.  The variable $T_0$ is the initial isothermal temperature, chosen to be $10^4$~K, and $\mu_0$ is the initial mean molecular weight, chosen to be 0.61.  A lower limit to the initial gas density of $10^{-20}$~\Msun\ pc$^{-3}$ is imposed affecting cells at high $h$.  We choose values of $\Sigma_0=10$ \Msun\ pc$^{-2}$ and $f_\rmn{g}=0.1$, which are similar to the gas surface density and gas fraction of the Milky Way's disc at the solar circle \citep{Flynn2006}.  The quantity $b_0$ is the isothermal scale height of the disc and is 100 pc for the models presented here.  The height containing half the mass in the box is related to $b_0$ by $h_{1/2} = 0.55b_0$.

Gas cells are part of a Voronoi mesh that is constructed around a set of `mesh-generating' points. The bulk of the mesh-generating points are distributed with uniform randomness in the horizontal $x$-$y$ plane and drawn randomly from a distribution function that follows Equation \eqref{eq:rho} in the $z$ direction.  The number of points placed in this fashion is $\Sigma_0 L^2 m_\rmn{t}^{-1}$, where $m_\rmn{t}$ is the target gas mass that the mesh refinement scheme will seek to maintain during the simulation.  For our fiducial model with $\Sigma_0=10$ \Msun\ pc$^{-2}$, $L = 1$ kpc, and $m_\rmn{t} = 10$ \Msun, this number is $10^6$ points.  These points are supplemented by a Cartesian background mesh that fills the entire volume and has a cell diameter of 50 pc.  The Cartesian mesh provides an initial minimum resolution for the gas in low-density regions at high $h$. Once the simulation begins, the Cartesian mesh deforms according the Voronoi mesh generation scheme of {\small AREPO}.

Once these mesh-generating points are placed, and before our science simulations are run, we run the initial conditions with {\small AREPO} in the `mesh-relax' mode.  This mode does not solve gravity, hydrodynamics, or any other physics module, but rather, the code only evolves the Voronoi mesh using the mesh-regularization algorithm \citep{Lloyd1982, Springel2010}.  This algorithm seeks to make cells more `round' by moving mesh generating points closer to their cells' centres of mass.  We typically run the initial mesh for approximately 100 timesteps in this mode with a slightly reduced maximum cell opening angle factor of $1.75$ (in our science runs, we return to the default {\small AREPO} value of $2.25$) until the mesh no longer changes.  After, gas densities are reassigned to the new relaxed mesh based on cells' new height $h$.  This step eliminates awkwardly shaped cells that can result from the initial random placement of mesh generating points.  It has the effect of regularizing the mesh in the midplane, but leaves the Cartesian mesh at large $h$ largely unchanged.

\subsection{Hydrodynamics and Magneto-hydrodynamics}
To solve the ideal hydrodynamics equations, we use the second-order accurate moving-mesh code {\small AREPO}  \citep{Springel2010, Pakmor2016a} with a thermal adiabatic index $\gamma$ of $5/3$.\footnote{Any gas in our simulations that is predominantly molecular is generally too cold to excite the rotational energy levels of H$_{2}$, and hence is well-described by an adiabatic index $\gamma = 5/3$.}  We have a target gas mass for cells of 10 \Msun\ and we impose a maximum timestep of 0.1 Myr.  In many of our simulations, ideal magnetohydrodynamics (MHD) is included with the Powell 8-wave scheme for divergence control \citep{Pakmor2013}.  When using MHD, an initial seed field that varies with height $h$ above the midplane is adopted with a strength of $10^{-7}\, {\rm  G} \times {\rm sech}^{4/3}(h/b_0)$, which scales the field as $\rho_0^{2/3}$ following Equation \eqref{eq:rho}. Its initial orientation is purely horizontal and parallel to the gas midplane, pointing in the $x$ direction.  This initial configuration is divergence free.

\subsubsection{Cooling and Chemistry}

Radiative cooling and chemistry are modelled following \citet{Smith2014} who use atomic and molecular hydrogen chemistry and cooling \citep{Glover&MacLow2007a,Glover&MacLow2007b} along with a simple treatment for CO chemistry \citep{Nelson&Langer1997,Glover&Clark2012} with the {\small AREPO} code to simulate the chemical evolution of gas in the disk of the MW.
Hydrogen is assumed to be present initially in fully atomic form, while carbon starts as C$^{+}$. The other species tracked in our chemical model (H$^{+}$, H$_{2}$ and CO) develop as the simulation progresses.
The abundances by number relative to hydrogen for carbon and oxygen are chosen to be $1.4 \times 10^{-4}$ and $3.2 \times 10^{-4}$ \citep{Sembach2000}. The ratio of dust to gas is taken to be 0.01.  The abundance of helium relative to hydrogen by number is 0.1.  A spatially uniform ultraviolet radiation field is applied throughout the volume that has a strength of 1.7 times the Habing field \citep{Habing1968}, similar to the MW's interstellar radiation field \citep{Draine1978}.  Gas self-shielding and dust shielding locally attenuate this field. We account for their effects using the TreeCol algorithm \citep{Clark2012} with a maximum shielding length of 30 pc.  Metal cooling of high-temperature gas assuming collisional ionization equilibrium is included with an assumed constant and spatially uniform gas metallicity of 1 \Zsun\ \citep{Gnat&Ferland2012,Walch2015}.  
Ionization of atomic and molecular hydrogen from CRs is also included with a spatially uniform field that we set to have a strength of $3 \times 10^{-17}$ s$^{-1}$ for H and twice that value for H$_2$ \citep{Smith2014}.  In order to have consistent ionization fields between runs done with and without CRs we choose to not use a spatially varying CR energy density field to vary these ionization rates.  By keeping ionization fields the same between all simulations, we can focus on the impact of CR pressure on the phase structure of the gas.

\begin{figure*}
\centering
\includegraphics[width=\textwidth]{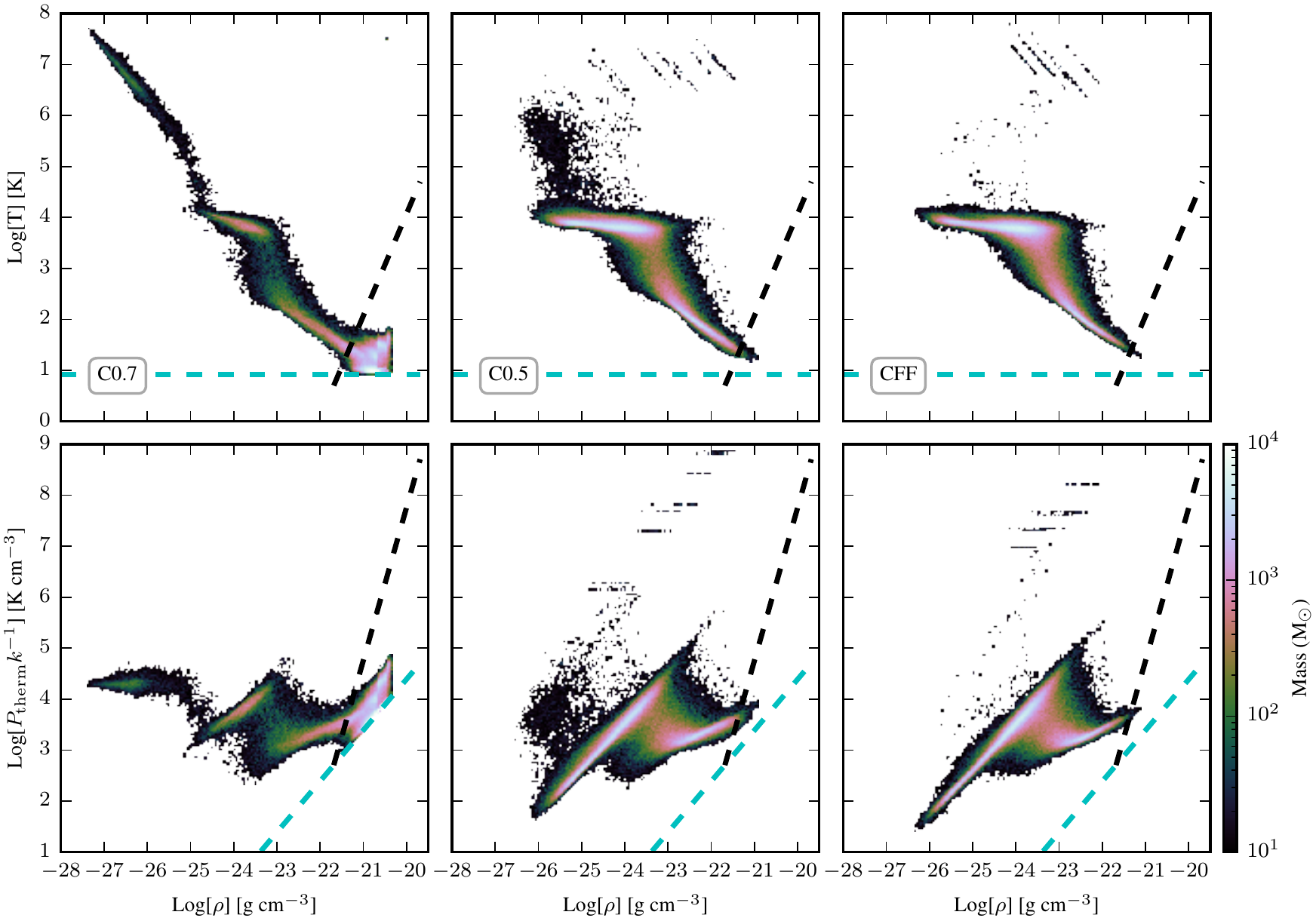}
\caption{Density, temperature, and pressure of gas within 1 kpc of the midplane after 100 Myr of evolution for three representative runs with magnetic fields and CRs.  Models with differing SN explosion placement are shown (see Table \ref{table}).  The fraction of SN explosions placed in dense gas increases from left to right (see Section \ref{sec:SNplacement} for details).  The intensity of the colour corresponds to the cumulative mass at each parameter-space coordinate: gas temperature vs. mass density (top) and thermal pressure vs. mass density (bottom).  The density where the Riemann pressure floor activates is shown with a black dashed line and the pressure and temperature limits from the minimum allowed specific thermal energy is shown with cyan dashed lines.}
\label{fig:phasespace}
\end{figure*}

\subsubsection{Thermal Pressure Support}

The low temperature cooling provided by molecular gas and the gas self-gravity that we employ results in the collapse of gas to very high densities.  Without a sink for dense gas or effective energetic feedback, a runaway collapse can occur.  To model this collapse, which in the real ISM would result in the creation of dense molecular cores and stars that have sizes below the resolution of these simulations, we limit the pressure used in the Riemann solver.  This approach has been used in AMR codes that place a limit on the finest mesh refinement level \citep[e.g.][]{Machacek2001}.  This approach was used in \citet{Simpson2016} where a minimum cell size was enforced.  Here we have altered our approach to one better suited to a Lagrangian code.  We begin with the assumption of a fixed mass resolution, and then compute the pressure of a Jeans stable cloud that can be resolved at this resolution.

We assume that a resolved cloud requires $N_{\rm{cloud}}$ gas cells.  For a fixed cell mass $m_\rmn{t}$, this cloud will have a mass of $M_
\rmn{J} = N_{\rm{cloud}} m_\rmn{t}$.  We estimate the mean density of a Jeans stable cloud of this mass with a mean temperature 10 times our chosen minimum allowed temperature (see below).  This density is
\begin{equation}
\label{eq:rhomax}
\rho_{\rm{max}} = \left(\frac{2\pi}{3N_{\rm{cloud}} m_\rmn{t}}\right)^2 \left(\frac{\pi \gamma (\gamma - 1) u_{\rm{min}}}{G}\right)^3.
\end{equation}
\noindent  The pressure of gas at this density and temperature is
\begin{equation}
\label{eq:Pmax}
P_{\rm{max}}= (\gamma -1) u_{\rm{min}} \rho_{\rm{max}},
\end{equation}
\noindent where $u_{\rm{min}}$ is the specific internal energy of gas at 10 times the minimum temperature, i.e.\ 50~K.  Once gas reaches the temperature floor, its pressure will scale linearly with density.  However, as the gas density rises above \rhomax, the mass resolution is insufficient to resolve a Jeans stable cloud at that mean density.  We therefore modify the fluxes computed in the Riemann solver to account for this issue.

We employ a pressure limit in the Riemann solver that follows a polytropic equation of state.  For gas densities of $\rho_i$ in cell $i$ a minimum limit is imposed for both the left and right states in the Riemann solver:
\begin{equation}
\label{eq:Pfloor}
P_{\rmn{floor},i}= P_{\rm{max}} \times \left(\frac{\rho_i}{\rho_{\rm{max}}}\right)^\alpha.
\end{equation}
\noindent The sound speed squared also has a floor of
\begin{equation}
\label{eq:cs}
c_{\rmn{s},i}^2 = \gamma \frac{P_{\rm{floor},i}}{\rho_i}.
\end{equation}

\begin{figure*}
\centering
\includegraphics[width=\textwidth]{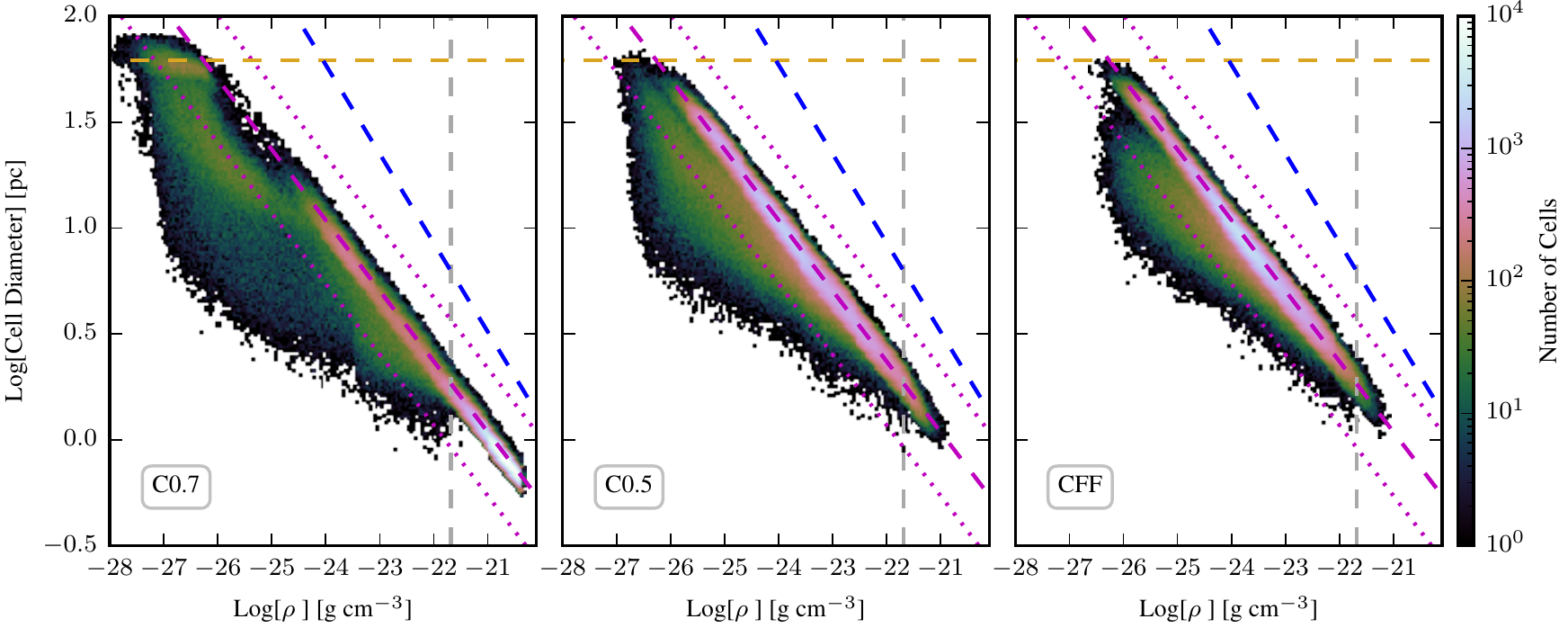}
\caption{
Resolution elements within 1 kpc of the midplane for three representative runs after 100 Myr of evolution that include magnetic fields and cosmic rays, and SN explosions.  
The `cell diameter' (defined as $(6 V_i/\pi)^{(1/3)}$ where $V_i$ is the cell volume) is plotted versus the cell gas density.
The color intensity indicates the number of cells at each cell diameter and density.  
Magenta lines trace constant cell mass at the target gas mass $m_\rmn{t}=10$ \Msun (dashed) and at $2 \times$ and $0.5 \times$ $m_\rmn{t}$ (dotted).  The diameter corresponding to the maximum cell volume (yellow dashed line) and the density where the pressure floor activates (grey dashed line) are also shown.  Twice the cooling radius of a SNR with an explosion energy of $10^{51}$ erg is shown as a function of background density \citep[e.g.][$\propto (\rho/m_p)^{-0.42}$, blue dashed line]{Draine}. }
\label{fig:resolution}
\end{figure*}

For $\alpha = 1$, the Riemann pressure floor would just be the cell's pressure, and the Riemann solver computes fluxes normally.  To limit the flux of mass into dense clouds, we adopt $\alpha > 1$.  The purpose of this study is not the detailed evolution of molecular clouds, we instead want to understand the overall phase structure of the ISM, including the hot and warm phases, where this numerical issue is not present, so modifying the pressure flows within dense clouds is an acceptable compromise.  We have adopted a value of $\alpha = 3$, which we found halted the collapse of gas to densities greatly exceeding \rhomax\ and did not affect the properties and evolution of gas at densities $\le \rho_{\rm{max}}$.  Figure \ref{fig:phasespace} shows the pressures of gas using this model.

For our fiducial resolution, the target cell mass is 10 \Msun\ and the minimum allowed temperature is 5 K.  If $N_{\rm{cloud}}$ is $10^3$, the Jeans mass $M_\rmn{J}$ is $10^4$ \Msun.  The maximum stable density of a cloud of this mass at ten times the minimum allowed temperature is $\rho_{\rm{max}} = 3.1$ \Msun\ pc$^{-3}$, or $2.1 \times 10^{-22}$ g cm$^{-3}$.  In cells greater than this density, our model adjusts pressure gradients to push outward against the strong force of self gravity that pulls gas inward.  The sound speed criterion is necessary to ensure time-stepping remains Courant limited in these regions.  

We have framed the motivation for this model in terms of the Jeans stability of a gravitating cloud by framing it in terms of the mass of the cloud ($N_{\rm cloud} m_{\rm t}$) and the specific energy of the cloud ($u_{\rm min}$).  However, the parameters $N_{\rm cloud}$ and $u_{\rm min}$ in Equation \eqref{eq:rhomax} are degenerate with each other and the only parameter we have introduced in this model that has actual utility in the code is \rhomax.  Ultimately, it is the value of \rhomax\ that is germane to the behaviour of the gas.  We have performed tests adjusting \rhomax\ at fixed $m_\rmn{t}$ in Appendix \ref{appendix}.  The evolution and properties of our models will be discussed in further sections, but Appendix \ref{appendix} shows that the pressure floor model adjusts the maximum gas density reached in the simulation while preserving the overall phase structure of the medium.

\subsubsection{Gravity}
Gravity is included from both a static analytic potential representing a stellar disc and from the gas self-gravity.  

The analytic stellar potential $\phi$ is a function of height $h$ and is related to the initial gas density and gas fraction via Poisson's equation: 
\begin{equation}
\label{eq:Poisson}
\nabla^2\phi = 4\pi G \rho_0(h) \times (f_\rmn{g}^{-1} - 1).
\end{equation}
\noindent The function $\rho_0(h)$ is the initial gas density given by Equation \eqref{eq:rho}.

Gas self-gravity is computed with a tree-based algorithm with an adaptive softening length for gas cells that is related to a cell's volume by $\propto V^{1/3}$ \citep{Springel2010}.  We impose a minimum softening length of
\begin{equation}
\label{eq:minsoft}
\epsilon_{\rm{min}}= \left(\frac{m_\rmn{t}}{\rho_{\rm{max}}}\right)^{1/3}
\end{equation}
\noindent For the fiducial model simulations the value of $\epsilon_{\rm{min}} = 1.48$ pc.  
The tree algorithm utilises an Ewald summation to capture long range gravitational forces at infinite distances across the periodic $x$ and $y$ boundaries \citep{GADGET4}.

\subsubsection{Mesh refinement and resolution}
Gas cells can be split or combined following a set of refinement and derefinement criteria that are dependent on the cells' masses, volumes, and Jeans lengths.  Cells are refined to maintain masses within a factor of two of the target cell mass $m_\rmn{t}$.  Cells are also refined in order to limit the volume ratio between adjacent cells to be no more than 10.  In cases where the cell's diameter is more than 4 times the Jeans length (following \citet{Truelove}) or where the maximum pressure difference among neighbouring cells is more than 10 times the cell's pressure, refinement is also applied.  This latter criterion is only applied in gas less dense than $\rho_{\rm{max}}$.  No cell is allowed to have a volume greater than $2.5 \times 10^5$ pc$^{3}$ (which corresponds to an effective radius of 31 pc) and no limit is set for the minimum allowed cell volume.  

The gas pressure floor and minimum softening halts the collapse of dense gas and truncates the distribution of cell sizes.  We define an average cell diameter as $(6 V_i/\pi)^{1/3}$.  Figure \ref{fig:resolution} shows the distribution of cell diameters for a few sample simulation outputs.  The cell radius at which the pressure floor becomes relevant is below 2~pc, so only a small fraction of the simulation volume is covered by these cells. However, these cells can contain a large amount of mass if numerous enough.  Other scales of the ISM are well resolved.  The scale height of the disc and the SN remnant (SNR) cooling radius at the volumetric average gas density are resolved by many cells unaffected by the pressure floor.  The cooling radius at the mass-weighted average density is also a factor of $\sim3$ greater than the median cell radius. 

Figure \ref{fig:phasespace} shows the phase space distribution that results from this Riemann pressure floor model.  The dense gas supply is heavily influenced by the SN placement model, discussed in future sections, but the behaviour of gas with pressure below the Riemann pressure floor (shown by the cyan line in Fig. \ref{fig:phasespace}) is limited by this model.  This gas consists of dense clouds as can be seen by eye in Fig. \ref{fig:images}.  We do not address the internal structure or evolution of these clouds as this polytropic model is not intended to capture the physics that regulates those properties.  We note that the deposition of feedback energy in these clouds will raise the pressure locally and thus make the pressure floor non-relevant for those affected cells.  Appendix \ref{appendix} describes in more detail how this model behaves with different model assumptions.

\subsection{Cosmic ray energy and transport}

CRs are modeled as a second fluid with an adiabatic index of $\gamma_{\rm{CR}} = 4/3$ as described by \citet{Pfrommer2017a}.  
This fluid tracks and evolves the energy density of CRs and couples the CR fluid to the thermal gas fluid via the CR pressure.
We additionally include CR transport in the form of anisotropic diffusion as described by \citet{Pakmor2016}.  
This model assumes a spatially and temporally constant diffusion coefficient $\kappa$ along the direction of the magnetic field in the cell and zero in all other directions.  
We use a fiducial value of $\kappa$ of $10^{28}$ cm$^2$~s$^{-1}$ (and vary it from $10^{27}$ to $10^{29}$ cm$^2$~s$^{-1}$ to address the uncertainties associated with this choice).  The CR energy density in our simulations evolves according to this equation:
\begin{equation}
\frac{\partial \varepsilon_{\rmn{CR}}}{\partial t} + \bm{\nabla} \bcdot [\varepsilon_{\rmn{CR}} \bm{v} - \kappa \bm{b}(\bm{b} \bcdot \bm{\nabla} \varepsilon_{\rmn{CR}})] = - P_{\rmn{CR}} \bm{\nabla} \bcdot \bm{v} - \Lambda_{\rmn{CR}} + \Gamma_{\rmn{CR}},
\label{eq:cr_transport}
\end{equation}
\noindent where $\varepsilon_{\rmn{CR}}$ is the energy density of CRs, $\bm{v}$ is the gas velocity, $\bm{b}$ is a unit vector in the direction of the local magnetic field, $P_{\rmn{CR}}$ is the CR pressure, $\Gamma_{\rmn{CR}}$ is CR energy injected in SNe, and $\Lambda_{\rmn{CR}}$ is the loss of CR energy from hadronic and Coulomb processes.

These Coulomb and hadronic processes are dependent on the energy spectrum of CRs.  
Unlike schemes that dynamically follow the CR energy spectrum on top of the MHD \citep{Girichidis2020,Girichidis2022}, the two-fluid model we use does not self-consistently track the spectrum of CRs, and therefore, we employ the assumption of an equilibrium spectrum that balances CR injection and loss \citep{Ensslin2007,Pfrommer2017a}.
We assume that all Coulomb losses and one-sixth of hadronic losses are thermalised, resulting in a transfer of CR energy to thermal energy.  
The remaining hadronic losses (5/6) are assumed to be directly lost to $\gamma$ rays and neutrinos, which are radiated away for the optically thin conditions of the ISM.  

For the adopted typical spectral index of shock-accelerated CRs of $\alpha_\rmn{inj}=2.2$ \citep{Haggerty2020,Caprioli2020}, the Coulomb loss rate is taken from \citep{Pfrommer2017a}:
\begin{equation}
\Lambda_{\rm{Coul}} = 2.78 \times 10^{-16} \left( \frac{n_\rmn{e}}{\rm{cm}^{-3}} \right) \left( \frac{\varepsilon_{\rmn{CR}}}{\rm{erg~cm}^{-3}} \right) \rm{erg~s}^{-1} \rm{cm}^{-3}
\end{equation}
\noindent and the hadronic loss rate by
\begin{equation}
\Lambda_{\rm{hadr}} = 7.44 \times 10^{-16} \left( \frac{n_\rmn{e}}{\rm{cm}^{-3}} \right) \left( \frac{\varepsilon_{\rmn{CR}}}{\rm{erg~cm}^{-3}} \right) \rm{erg~s}^{-1} \rm{cm}^{-3}.
\end{equation}
\noindent These rates are used during the running of the simulation to provide sources and sinks to the CR and thermal fluids as we have described.

 \subsection{SN injection models}
\label{SNplacement}
SN explosions are modelled as discrete events depositing $10^{51}$~erg of energy in the simulation when they occur.  They are deposited in the gas by injecting energy into the 32 closest cells surrounding the site selected for the explosion.  This site can either be the centre of a gas cell with a desired property or a point in space selected by other criteria.  

The SN energy that is injected is split between thermal energy and CR energy.  We choose to inject 10 per cent of the SN energy in CRs and 90 per cent in thermal energy.  This value is consistent with previous models \citep[e.g.][]{Simpson2016} and some observations \citep[e.g.][]{Morlino2012} but may be on the high side of acceptable values if magnetic obliquity-dependent acceleration efficiencies are integrated around the entire supernova remnant \citep[][]{Caprioli2014,Pais2018}.  Both types of energy are injected proportionally to a cell's volume in the SN region.  In models without CRs, all $10^{51}$ erg of energy is deposited in thermal energy.  The chemical composition of cells are not adjusted during the injection.

We use two methods for SN explosion placement.  The first method (called FF) was used in \citet{Simpson2016} and selects SN explosion locations based on a local SFR computed from the local dynamical free-fall time:
\begin{equation}
{\rm sfr}^{\rm ff}_i = \epsilon \frac{m_i}{t_{{\rm ff},i}},
\label{eq:sfr}
\end{equation}
\noindent where $m_i$ is the mass of cell $i$, $\epsilon$ is a star formation efficiency (we adopt 1 per cent), and $t_{{\rm ff},i}$ is the dynamical free-fall time within the gas cell:  
\begin{equation}
t_{{\rm ff},i} = \sqrt{\frac{3\pi}{32 G \rho_{\rmn{b},i}}}.
\label{eq:ff}
\end{equation}
The dynamical free-fall time is a function of the baryon density with the cell $\rho_{\rmn{b},i}$, which is the sum of the gas density $\rho_i$ and the stellar mass density given by the background potential profile.

We assume the rate of SN explosions in each cell is related to ${\rm sfr}^{\rmn{ff}}_i$ by the factor of 1.8 explosions per 100 \Msun\ of newly-formed stars. The probability of a supernova explosion at the position of cell $i$ in time step $\Delta t$ is therefore
\begin{equation}
\label{eq:prob}
p^{\rm ff}_i = {\rm sfr}^{\rmn{ff}}_i \times \frac{1.8 \mathrm{\ SNe}}{100 \mathrm{\ M}_{\odot}} 
\times \Delta t,
\end{equation}
where we have assumed that $\Delta t \ll t_{{\rm ff},i}$.  Indeed, this is the case as seen in Fig. \ref{fig:timescales} in Appendix \ref{appendix_timescales} where we see that $t_\rmn{ff}$ is well above the maximum timestep of 0.1 Myr.  In dense gas where $t_{\rm ff}$ is shorter, the time steps are typically much less than 0.1 Myr in order that the evolution of the thermal gas is Courant limited.

The second method of SN placement (called MIX) adds a random-placement component to the FF model.  This component reflects SN explosions that do not occur in their natal molecular cloud, such as those arising from run-away massive stars, or that occur in star-forming environments where dense gas has already been cleared away by prior SN explosions, photoionisation or winds from Wolf-Rayet stars. For this method, we select a fraction of randomly placed SN explosions, $f_{\rm rand}$, and divide the total number of SN explosions between the two modes.  The total number of SN explosions is fixed according to the empirical Kennicutt-Schmidt relation \citep{Kennicutt1998} that gives the surface density of star formation $\dot{\Sigma}_{\rm KS} = 2.5 \times 10^{-4} \Sigma_\rmn{g}^{1.4}$ [\Msun\ kpc$^{-2}$ yr$^{-1}$], as a function of gas surface density $\Sigma_\rmn{g}$ [\Msun\ pc$^{-2}$].  The number of randomly placed SN explosions in time step $\Delta t$ is
\begin{equation}
N_{\rm rand} = \dot{\Sigma}_{\rm KS} \times L^2_{\rm box} \times \frac{1.8 \mathrm{\ SNe}}{100 \mathrm{\ M}_{\odot}} \times f_{\rm rand} \Delta t.
\end{equation}

\noindent SN explosions injected in this mode are placed with uniform randomness in the horizontal $x-y$ plane.  Their vertical $z$ position is drawn randomly from a distribution function that follows Equation \eqref{eq:rho}, so that events are concentrated in the midplane.  These randomly placed SN explosions therefore represent a population of core-collapse events from massive stars, however, explosions from Type Ia SN would also have no spatial correlation with dense gas, but would also not depend on the Kennicutt-Schmidt relation.  

The remaining SN explosions in MIX are placed according to local gas properties in a manner similar to FF.  Each cell is given a local SFR for this mode:
\begin{equation}
{\rm sfr}^{\rm mix}_i = {\rm sfr}^{\rm ff}_i \times \frac{\dot{\Sigma}_{\rm KS} L^2_{\rm box}}{\sum{{\rm sfr}^{\rm ff}_i }} \times (1 - f_{\rm rand})
\end{equation}
\noindent and the probability $p^{\rm mix}_i$ of a SN explosion occurring at the position of cell $i$ is computed in the same way as Equation \eqref{eq:prob}.  

By normalizing the local SFR in this way, the MIX model fixes the global SFR rate to the Kennicutt-Schmidt rate, while maintaining the relative spatial differences in the SFR used to place SNR in the non-random mode due to variations in gas density.  In the MIX model, the global SFR is adjusted over time as gas flows out of the simulation box at the outflow boundaries and $\Sigma_\rmn{g}$ decreases.  In the FF model, the global SFR varies due to changes in local gas gas properties, the effect of which is reflected in the global rate because the global rate is the sum of all the local SFRs.  

The net result of increasing \frand\ is that more SN energy will be coupled to low density gas on injection.  Given our Lagrangian mesh refinement scheme, this means the injection zones in models with larger \frand, while always comprising 32 cells and containing approximately 320 \Msun of gas, will occupy a larger volume.  However, the mesh resolution requirement for capturing the cooling of SN energy deposited on a mesh decreases with decreasing background density \citep[e.g.][]{Simpson2015}.  Also, the cooling time in low density gas in runs where it dominates the volume is typically longer than the simulation time (see Fig. \ref{fig:timescales}).

For consistency with the assumed SFR we compute locally, we also employ a step of draining star forming gas from cells and re-injecting it onto the mesh to account for gas loss from star formation.  This is done instantaneously within a time step, but a very small amount of gas is cycled in this way, reflecting the low efficiency of star formation on the scales we consider here.  At each time step and in each cell $i$, we estimate the amount of gas mass that goes into stars as ${\rm sfr}_i \times \Delta t$.  For the FF model, ${\rm sfr}_i$ is simply ${\rm sfr}_i^\rmn{ff}$ and for the MIX model, it is ${\rm sfr}_i^{\rm{mix}}/(1- f_{\rm{rand}})$.  If this mass is less than the mass of the cell, it is removed.  Once all star forming mass is drained, the total amount of gas is redistributed on the mesh.  In this redistribution phase, each cell receives an amount of gas proportional to $\rho_0(h_i) (1/f_{\rm{g}} - 1) V_i$, which is the mass of stars implied by our assumed potential, gas fraction, and the cell volume.  When mass is drained the specific CR energy is kept constant, resulting in a loss of CR energy.  When gas is injected, the specific CR energy is changed to keep the CR energy constant.  Thus, this cycling of star-forming gas is a net sink for CRs but not a source of CR heating.  The magnetic energy is kept fixed during both steps.

\begin{figure}
\centering
\includegraphics[width=\columnwidth]{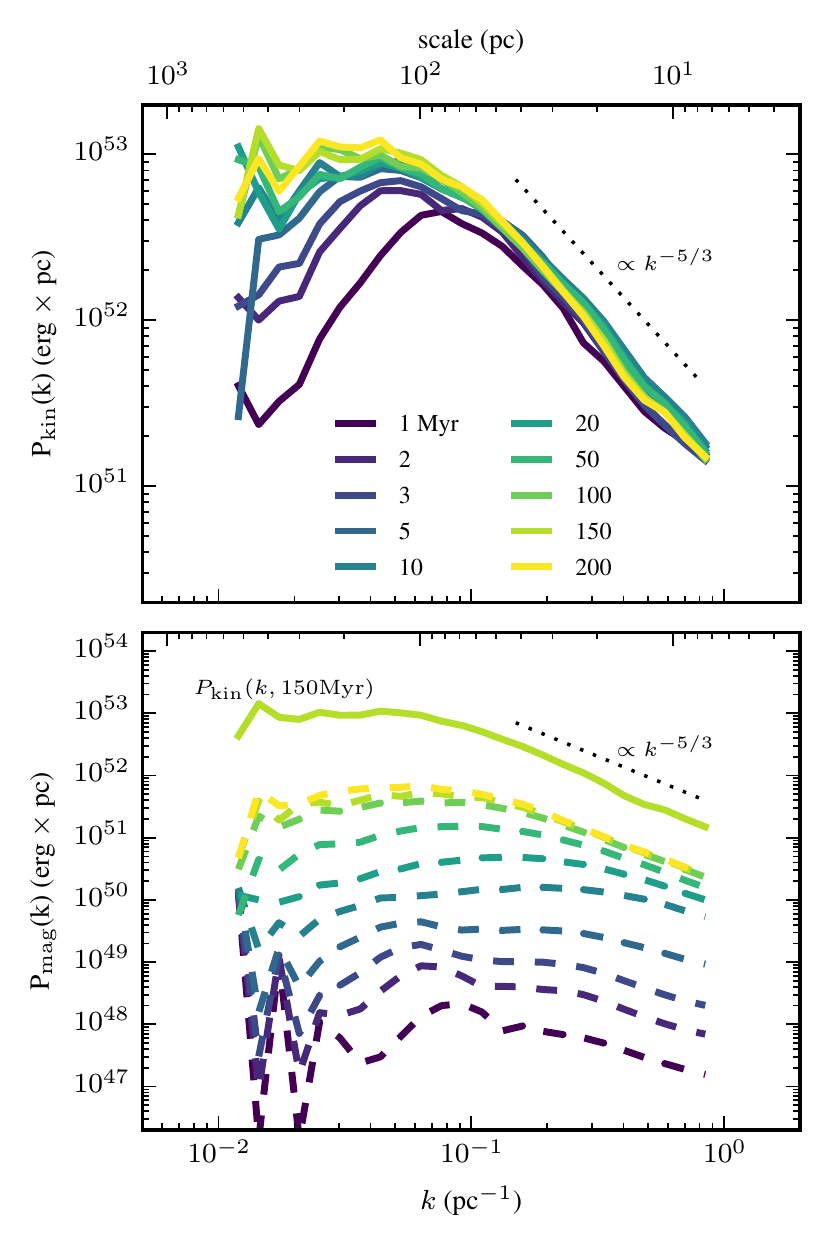}
\caption{Top: Power spectra of kinetic energy within 250 pc of the box midplane at different times ranging from 1 to 200 Myr in simulation C0.5, which contains CRs and magnetic fields.  The Kolmogorov turbulence slope of $-5/3$ is indicated by a dotted line.  Bottom: Power spectra of magnetic energy in the same volume shown with dashed lines.  For comparison, the power spectrum of kinetic energy at 150 Myr from the top panel is shown.  The power-law slope of $-5/3$ is also indicated.}
\label{fig:turbulence}
\end{figure}

Our SN injection model does not directly inject kinetic or magnetic energy, but the pressure forces caused by the injection of thermal energy induces gas motions that results in a turbulent medium, which in turn, amplifies the seed field.  Fig.~\ref{fig:turbulence} shows the development of the turbulent spectrum of kinetic energy that has a Kolmogorov-like slope from scales of pcs to tens of pcs and saturates very early in the simulation.  The spectrum of magnetic energy (also shown) grows more slowly but also reaches saturation by 100 Myr.

\section{How CRs mediate the effect of SN explosions on ISM structure}
\label{sec:mainresults}
In this section we focus on how the structure of the ISM changes when different models for SN placement are applied.  The models for SN placement that we test vary the number of SN explosions placed in high-density gas and the number of remnants placed randomly in the midplane.  We explore how varying the fraction of SN energy coupled to low density gas (in models with high numbers of randomly placed SN explosions) increases the volume filling factor of hot gas ($\ge 2 \times 10^4$ K) and we will then explore how the presence of CR pressure modifies this effect.
Table~\ref{table} summarises the simulations run to model these effects.

\begin{table}
\caption{List of simulations.}
\label{table}
\begin{tabular}{lclll}
\hline
Simulation & $f_{\rm{rand}}$ & MHD & CR & $\kappa$ (cm$^2$s$^{-1}$)\\
  \hline
 HFF & - & no & no & - \\
 CFF & - & yes & yes & $10^{28}$ \\
  \\
 H0.1 - H1.0$^*$ & 0.1 - 1.0 & no & no & -\\
 C0.1 - C1.0$^*$ & 0.1 - 1.0 & yes & yes & $10^{28}$\\
 \\
 CFF-$\kappa$27-29$^{**}$ & - & yes & yes & $10^{27}$ - $10^{29}$\\
 C0.5-$\kappa$27-29$^{**}$ & 0.5 & yes & yes & $10^{27}$ - $10^{29}$\\
 CFF-adv$^\dagger$ & - & yes & yes & 0\\
 C0.5-adv$^\dagger$ & 0.5 & yes & yes & 0\\ 
 \hline     

 \end{tabular}
 
 \raggedright $^*$These simulations vary \frand\ from a value of 0.1 to 1.0 as indicated in the simulation label.
 
 \raggedright $^{**}$These simulations vary the CR diffusivity $\kappa$ from a value of $10^{27}$ cm$^{2}$s$^{-1}$ to $10^{29}$ cm$^{2}$s$^{-1}$.  The value of Log($\kappa$) is indicated in the simulation label.
 
 \raggedright $^\dagger$These simulations include the CR fluid, but only allow it to advect with the thermal gas; the diffusion term from Equation \eqref{eq:cr_transport} is not included (i.e. $\kappa = 0$).

\end{table}

\subsection{Varying the placement of SN explosions}
\label{sec:SNplacement}
\begin{figure*}
\centering
\includegraphics[width=\textwidth]{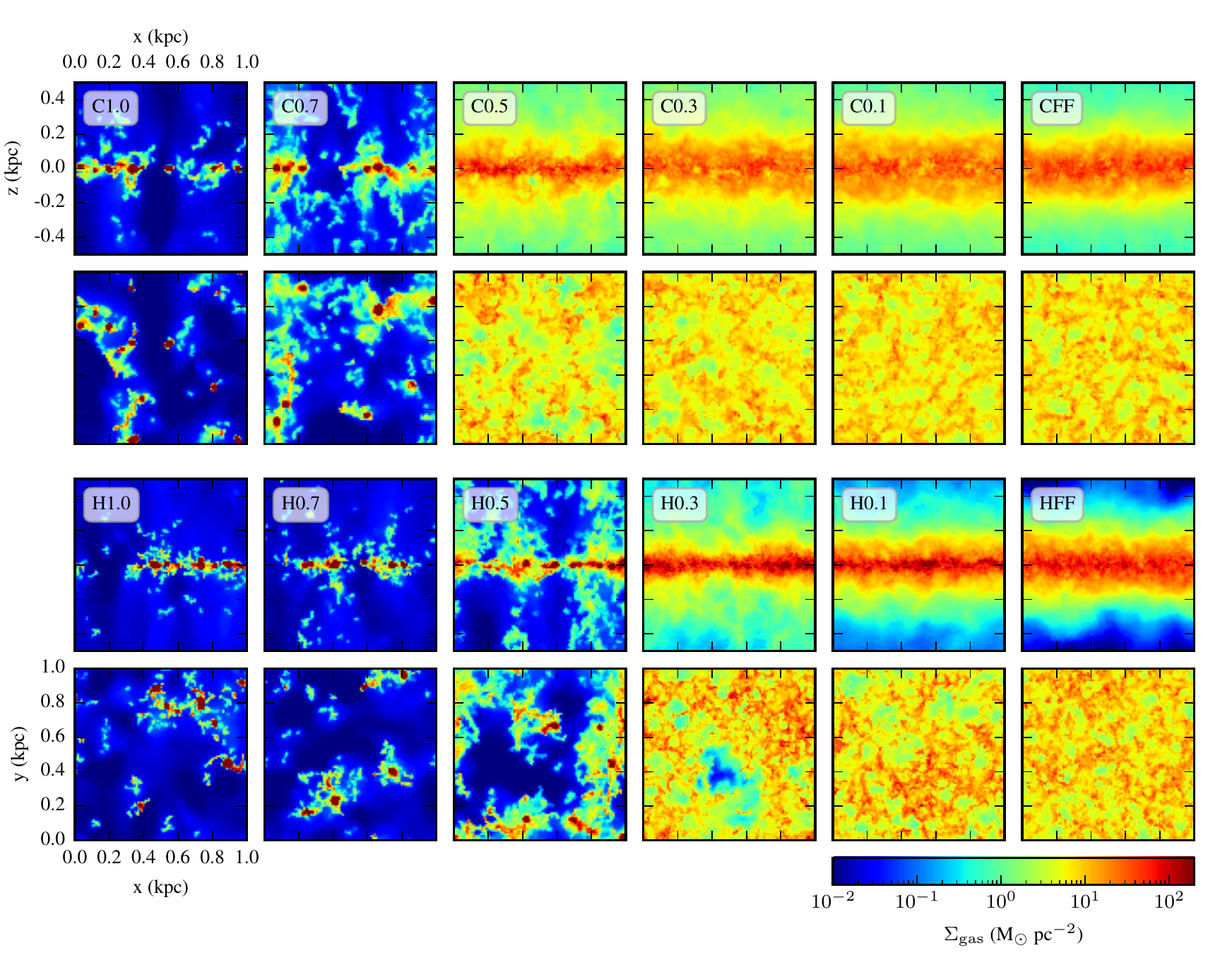}
\caption{Projections of gas number density of the central square kiloparsec of models with CRs (top two rows) and without CRs (bottom two rows)  In each pair of rows, two orthogonal projections are shown: edge-on (top) and face-on (bottom).  From left to right, simulations with varying SNR injection models are shown, and are in order MIX models with \frand\ = 1, 0.7, 0.5, 0.3, and 0.1, and then the FF model.}
\label{fig:images}
\end{figure*}

Figure \ref{fig:images} shows a snapshot of the state of the gas after 100 Myr of evolution in a representative sample of our models with different SN explosion placement.  It shows the central kpc of each simulation volume that includes the multiphase midplane gas.  Figure \ref{fig:images} shows that gas in the midplane becomes more porous as models include higher fractions of randomly placed SN explosions.  In models with predominantly random SN explosions, a `thermal runaway' occurs in the midplane, where the coupling between SN energy and dense gas is inefficient and where gas self-gravity and radiative cooling cause clouds to become denser and denser.  Subsequent explosions tend to occur in hot, rarefied gas as the hot phase occupies most of the volume \citep[e.g.,][]{McKee1977}.  This heats gas further, and the end result is that the volume fraction filled by gas with temperatures in excess of $2 \times 10^5$ K exceeds 90 per cent.  In models with high $f_{\rm{rand}}$, the midplane mass becomes dominated by a population of dense, self-gravitating clouds.  The size of these clouds is regulated by the Riemann pressure floor that acts to keep them Jeans stable, i.e.\ prevents them from collapsing further.  

In models with low $f_{\rm{rand}}$, SN energy couples to dense gas more efficiently because of the placement of more SN explosions in dense gas.  This `peak driving' prevents gas from collapsing to dense clouds, consistent with previous studies \citep[e.g.][]{Walch2015}.  In these models, `warm gas' (with temperatures between 300 K and $2 \times 10^5$ K) dominates the volume.

\begin{figure*}
\centering
\includegraphics[width=\textwidth]{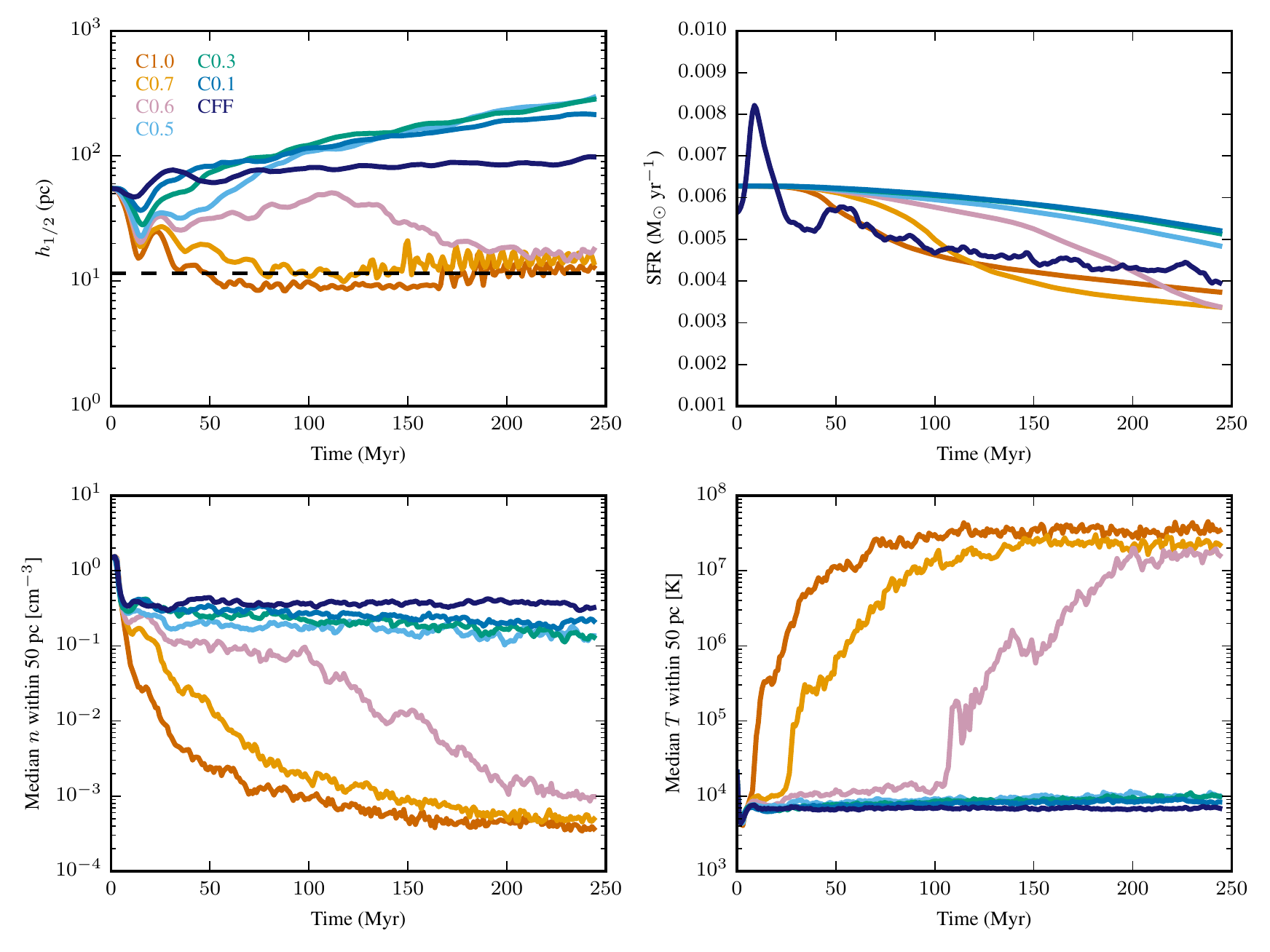}
\caption{Evolution of midplane properties for simulations with CRs.  Top left: Height containing half the original gas mass within the box.  The minimum radius of a Jeans stable cloud assumed by the pressure floor model is shown by a dashed line.  Top right: The total star formation rate of the box, which seeds the SNR injection rate.  Bottom left: The volume-weighted median gas number density within 50 pc of the midplane.  We compute this median value with respect to cell volume, i.e. this is the density at which half the volume is at lower densities and half at higher.  Bottom right: The volume-weighted median gas temperature within 50 pc of the midplane.  We adopt the same volume-weighted median definition as in the number density panel.  }
\label{fig:mix_evolution}
\end{figure*}

Figure \ref{fig:mix_evolution} shows how this process proceeds over time through the evolution of midplane properties.  In the MIX models, the global SFR is fixed to the Kennicutt-Schmidt rate of the box.  All the MIX models begin with the same SFR, but as time goes on, and as gas is lost from the box through outflows (the topic of a companion paper), the SFR drops because the gas surface density drops.  

In the FF model, the SFR is the sum of all the local rates given by Eq. \eqref{eq:sfr}, i.e. determined only by the local gas density and free-fall time, instead of being fixed to a global rate.  This gives an initial rate close to the Kennicut-Schmidt rate, but due to the initial radiative cooling of the midplane gas, a modest initial starburst occurs that gives a 50 per cent enhancement in the SFR.  By 50 Myr, this burst is arrested and an equilibrium state is reached in the FF model that balances radiative cooling, SN energy injection, and mass loss through outflows, resulting in a near constant SFR and median midplane gas properties.  The volume-weighted median midplane gas density and temperature (defined as a the density and temperature that bisect the volume-weighted distributions of these quantities), are essentially constant (0.3 cm$^{-3}$ and 7000 K), and the height of the central gas layer (defined as the height containing half the original gas mass) is also essentially constant in FF.

In the MIX models, the midplane evolution proceeds differently and depends strongly on $f_{\rm{rand}}$.  There appear to be two attractor solutions to which MIX models will evolve: one is the thermal runaway solution, which occurs when $f_{\rm{rand}} \ge 0.6$ and the other is the peak driving solution, which occurs when $f_{\rm{rand}} \le 0.5$.  In the thermal runaway state, midplane gas evolves to gas densities below $10^{-3}$ cm$^{-3}$ and temperatures in excess of $10^7$ K.  

The vertical distribution of gas also bifurcates into two states.  With large $f_{\rm{rand}}$ the mass in the midplane collapses to pressure-floor supported clumps as seen in Fig.~\ref{fig:images} that then results in a gas layer with a height given by the ratio of the Jeans mass we resolve ($M_\rmn{J}$) and the density ($\rho_{\rm{max}}$) where the pressure floor kicks in: 
\begin{align}
    h_{\rm{min}} =  \left(\frac{3}{4\pi} \frac{M_\rmn{J}}{\rho_{\rm{max}}}\right)^{1/3} \sim 12~\rmn{pc}.
\end{align}
With a small $f_{\rm{rand}}$, the height of the gas layer extends above 100~pc.  In the MIX models with a small $f_{\rm{rand}}$, the height of the gas layer does not reach an equilibrium, unlike FF, because the global SFR is larger and does not respond to instantaneous changes in the midplane; it adjusts downward only when gas is lost through the outflow boundaries.  This results in a constant puffing up of the midplane gas layer.  

The closest model to an `intermediate' state between the two attractor solutions is the $f_{\rm{rand}} = 0.6$ model (pink lines in Figure \ref{fig:mix_evolution}).  It has a slower rise in its gas scale height in the first 100 Myr and a midplane gas density close to the peak driving runs at 50 Myr.  However, this state is not stable, and after 100 Myrs, it evolves to the thermal runaway case.

\begin{figure}
\centering
\includegraphics[width=\columnwidth]{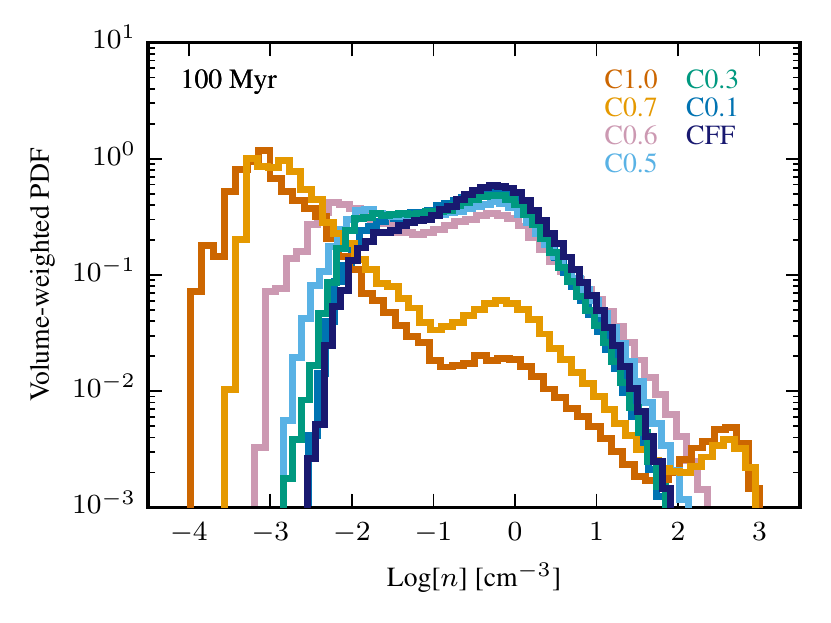}
\includegraphics[width=\columnwidth]{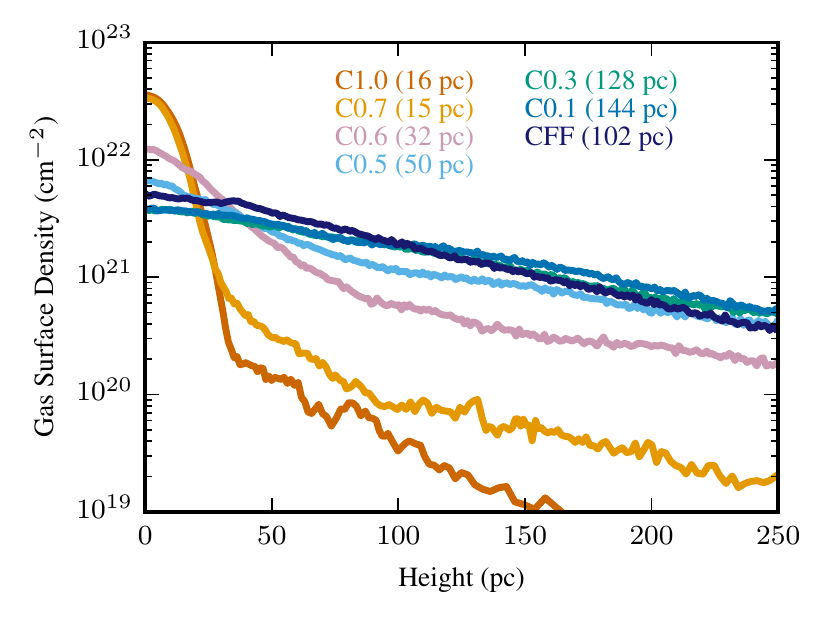}
\caption{Top: Volume-weighted probability density functions (PDFs) of gas density within 50 pc of the mid-plane for simulations with cosmic ray diffusion after 100 Myr of evolution.  The PDFs shown are of the logarithms of the gas number density, Log($n$).  Bottom: Surface density of gas in the vertical direction for models with cosmic ray diffusion after 100 Myr of evolution.  The simulation volume above and below the mid-plane are combined and the mid-plane is at height 0.  The height where the gas surface density drops by 1/e from the peak value at the mid-plane is indicated for each run.}
\label{fig:profiles}
\end{figure}

Figure \ref{fig:profiles} shows the differences in the distributions of midplane gas between the peak-driving and thermal run-away states.  The distribution in Log($n$) ($n$ is the gas number density) follows a skewed log-normal distribution in the peak driving case, with the skewness extending to the low density side.  The distribution in the thermal runaway case follows approximately a power law in Log($n$), with three bumps: one at densities above $\rho_{\rm{max}}$ due to pressure-floor supported gas clumps; one at Log($n$)~$ = 0$, where the peak driving distributions peak; and one at Log($n$)~$ = -3$ where the distribution turns over.  We note that the volume-weighted average of $n$ and the volume-weighted average of Log($n$) occur at different $n$, with the average in Log($n$) occurring at lower $n$.  

Figure \ref{fig:profiles} also shows the vertical distribution of gas density.  Simulations in the thermal runaway case have a rapid fall off in density beyond 11 pc, approximately the radius of a Jeans stable cloud at the Riemann pressure floor.  For simulations in the peak driving case, the gas density falls off much more gradually.  The typical e-folding height of gas (the height where the gas surface density drops by 1/e), is less than 20 pc in the first case and between 50 and 150 pc in the second case.  We note that our simulations do not include a circumgalactic medium (CGM) component that the disc medium can interact with, which would likely impact these vertical profiles at larger heights.

\subsection{Midplane energy partition}
\label{sec:energy}
\begin{figure*}
\centering
\includegraphics[width=\textwidth]{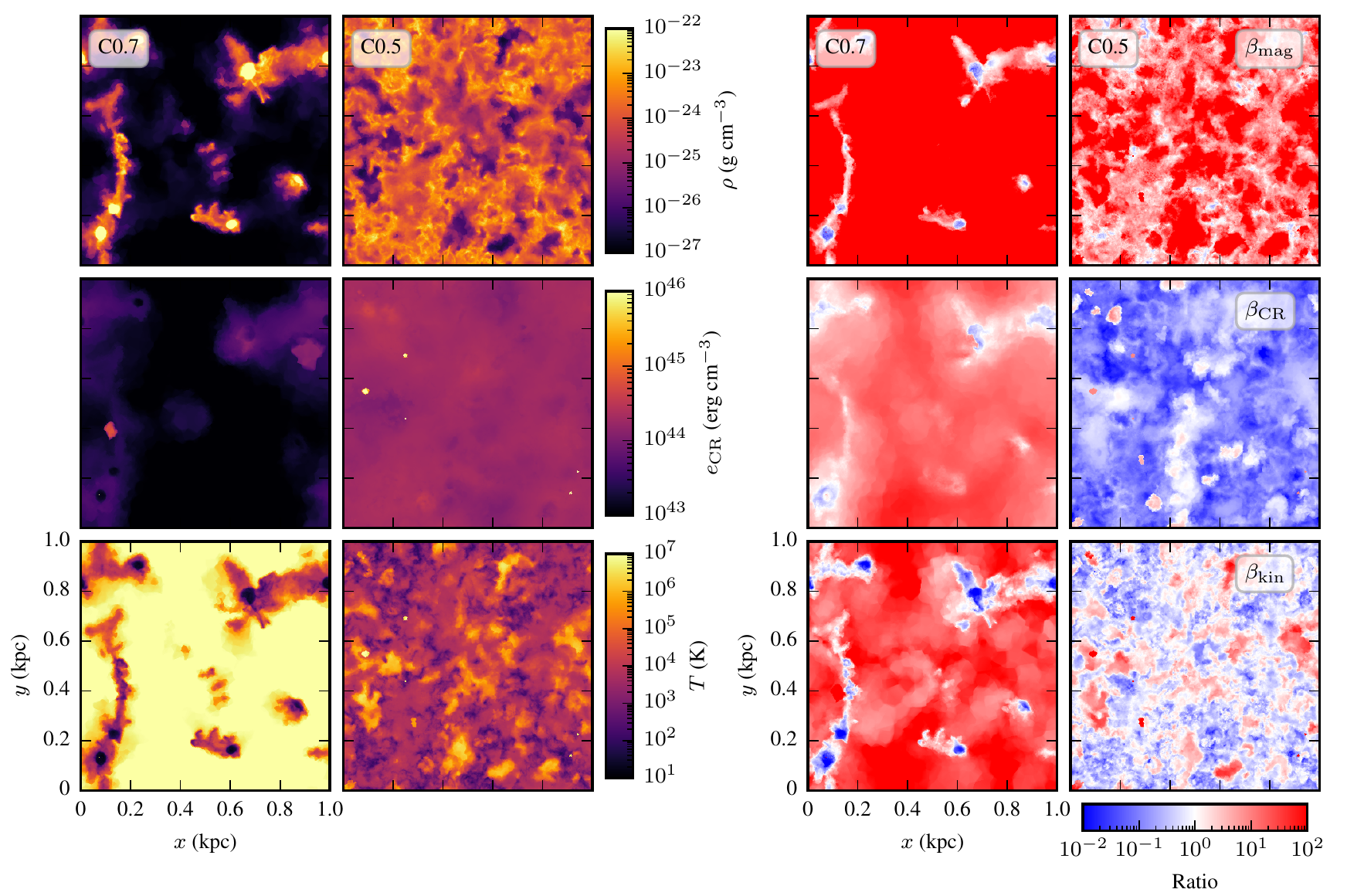}
\caption{Thin projections of midplane gas properties in the C0.7 and C0.5 models after 100 Myr of evolution.  The central 100 pc of the box are projected along the vertical axis showing a face-on view and the full with of the box (1 kpc $\times$ 1 kpc) is shown. Left: the panels show, from top to bottom, the gas density; the CR energy density weighted by the gas density; and the gas temperature weighted by the gas density.  Right: these panels show the ratio of energy column density maps for different energetic components.  From top to bottom, we show thermal energy density to magnetic energy density ($\beta_{\rm{mag}}$); the thermal energy density to CR energy density ($\beta_{\rm{CR}}$); and the thermal energy density to kinetic energy density ($\beta_{\rm{kin}}$).  White indicates regions of equal energy density and red thermally dominated regions.}
\label{fig:energyimages}
\end{figure*}

The phase structure differences we have described correlate with different energetic balances within the ISM.  Figure \ref{fig:energyimages} shows the spatial distribution of CR energy and temperature, along with the ratio of thermal energy to the three other energies in the simulation, magnetic, CR, and kinetic.  We define these ratios as $\beta_{\rm{mag}}$, $\beta_{\rm{CR}}$, and $\beta_{\rm{kin}}$:
\begin{equation}
    \beta_\mathrm{kin} = \frac{\varepsilon_{\mathrm{therm}}}{\frac{1}{2} \rho v^2},
\end{equation}
\begin{equation}
    \beta_\mathrm{CR} = \frac{\varepsilon_\mathrm{therm}}{\varepsilon_\mathrm{CR}},
\end{equation}
\begin{equation}
    \beta_\mathrm{mag} = \frac{\varepsilon_\mathrm{therm}}{B^2/(8\pi)},
\end{equation}
\noindent where $\varepsilon_\mathrm{therm}$ is the thermal energy density.  A $\beta$ value of 1 indicates a cell with equal amounts of thermal energy and the non-thermal energy under consideration.

In all models, the CR energy density is quite smooth throughout the midplane, with the exception of the location of very young SN explosions.  The thermal gas energy varies more, so the spatial variation of $\beta_{\rm{CR}}$ is set primarily by the thermal energy.  In the peak driving runs, most of the volume is dominated by CR energy and only in regions with young SNRs does thermal energy dominate due to our prescribed injection ratio of CRs.  In thermal runaway runs, most of the volume is dominated by thermal energy and CRs only dominate in cold, dense clouds.

The overall energy density of CRs does vary between high and low \frand\ runs, with a higher CR energy density level found in the midplane in the peak-driving runs with lower \frand.  With most of the volume of the peak-driving runs being filled with warm ($\sim10^4$ K) gas rather than hot ($\sim 10^7$ K) gas (as in the thermal-runaway case), on large scales the CR energy dominates the volume in the peak-driving case, unlike the thermal-runaway case.

High magnetic energy density is largely confined, in both types of model, to only the densest gas and is therefore subdominant in most of the midplane volume.  In models with large peak driving, the ratio of thermal to turbulent energy can vary on quite short length scales. In models with large random driving, the thermal energy is so great and the gas density so low so that the thermal energy exceeds the turbulent energy.

In the thermal-runaway case (e.g.\ C0.7), the thermal energy dominates on large scales and it is only in the dense clouds where magnetic and kinetic energies can dominate.  These clouds are strongly self-gravitating, but are subject to the pressure limiter, so their internal properties should be regarded in that light.  In the peak-driving case (e.g.\ C0.5), the kinetic and thermal energies are nearly in equipartition on large scales, but have variations on smaller scales following variations in the local thermal temperature and density which span a smaller dynamic range.

In all runs, the magnetic $\beta_{\rm{mag}}$ is large ($> 10$) in most of the volume.  Pockets where it is less than unity correlate with areas of dense gas, however in peak-driving runs (e.g.\ C0.5), there is a larger fraction of the volume with $\beta_{\rm{mag}} \sim 1$, found in $\sim 1000$ K gas.

\begin{figure}
\centering
\includegraphics[width=\columnwidth]{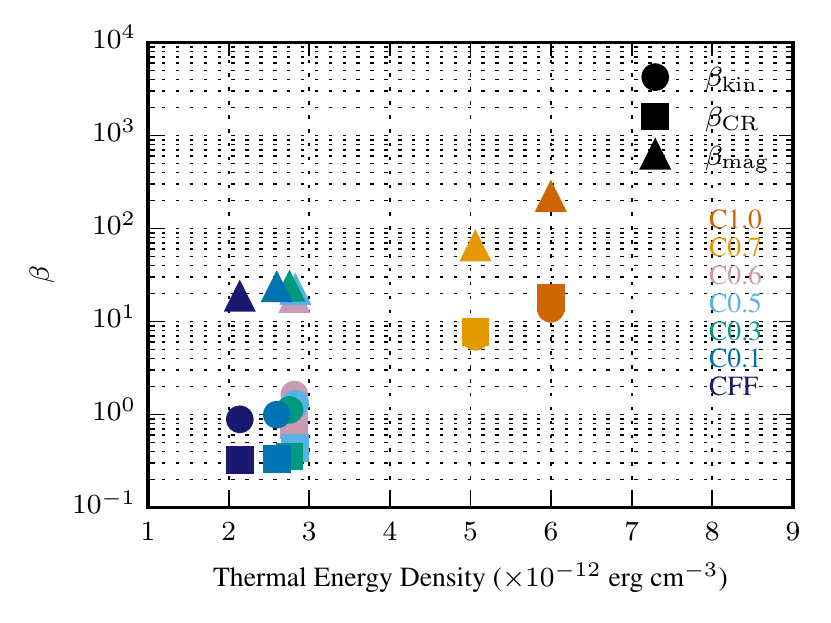}
\caption{Energy ratios $\beta_{\rm{kin}}$, $\beta_{\rm{CR}}$, and $\beta_{\rm{mag}}$ of gas within 50 pc of the midplane.  Quantities are averaged from snapshots between 95 and 104 Myr.  The thermal energy density is defined as the sum of all cells' thermal energy divided by the total volume of the midplane region (1 kpc $\times$ 1 kpc $\times$ 100 pc).  The ratios shown are the thermal-to-kinetic energy ($\beta_{\rm{kin}}$, circles), the thermal-to-CR energy (squares), and the thermal-to-magnetic energy (triangles).}
\label{fig:edensities}
\end{figure}

The trends on large scales are summarised in Fig.~\ref{fig:edensities}.
Models of the ISM at the solar circle have long indicated that the different pressures in the ISM, thermal, kinetic, CR, and magnetic, are nearly in equipartition at the midplane \citep{BoularesCox1990}.
The thermal-runaway models have larger densities of thermal energy in the midplane; however, the other energetic components of the ISM are not equivalently elevated, so that no other energy component is more than 10 per cent of the thermal energy density.  This contrasts with the peak-driving runs in which the kinetic energy density is within a factor of a few of the thermal energy density and the CR energy density is typically a factor of 2--3 higher than the thermal energy density.  The magnetic energy density is below 10 per cent of the thermal energy consistent with the saturation strength expected for a small-scale turbulent dynamo. By construction we lack cosmological \citep{Marinacci2016,Pakmor2017,Pfrommer2021} and disc amplification mechanisms \citep{Shukurov2006} that could amplify the magnetic field beyond this strength.

\subsection{Evolution of the CR energy}
\label{sec:energy_evo}

The $\beta_{\rm{CR}}$ parameter is determined by the evolution of the thermal gas, but also by the evolution of the CR fluid, which Fig.~\ref{fig:energyimages} shows does not have the same dependence on the phase structure of the gas as the thermal energy does.
There are several source and sink terms for the CR energy (see Eq. \ref{eq:cr_transport}) in our simulations.  There is the injection of CR energy through our SN feedback model model ($\Gamma_{\rmn{CR}}$); adiabatic changes of the CR fluid due to gas motions ($-P_{\rmn{CR}} \bm{\nabla} \bcdot \bm{v}$); CR cooling from both hadronic and Coulomb losses ($\Lambda_{\rmn{CR}}$); and losses due to star formation that arise when we remove star-forming mass from a cell and keep the specific CR energy constant.  In addition, CR energy can be lost through the outflow boundary, either advectively or diffusively.

\begin{figure*}
\centering
\includegraphics[width=\textwidth]{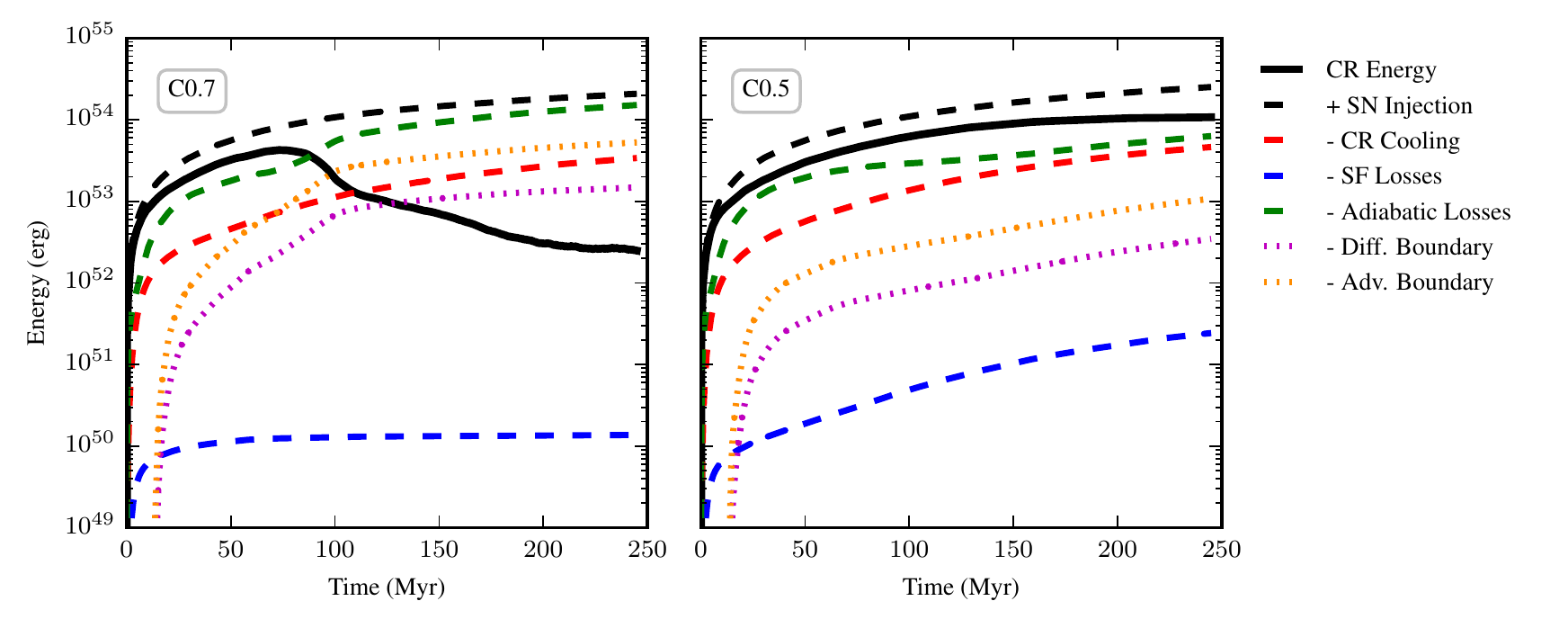}
\caption{Comparison of the CR energy evolution for the C0.7 run (left) which undergoes a thermal runaway and the C0.5 run (right) which does not and has warm ($\sim 10^4$ K) gas filling most of the midplane volume.  The quantities shown are summed over the entire simulation domain.  The instantaneous CR energy is shown (solid black line) and source and loss quantites are shown as cumulative sums over time.  The source term is the injection of CRs from SNe (black dashed).  The loss terms shown are CR cooling from hadronic and Coulomb processes (red dashed); losses from the removal of star-forming gas from the mesh (blue dashed); adiabatic losses (green dashed); adiabatic losses over the outflow boundary (orange dotted); and diffusive losses from the outflow boundary cells (magenta dotted).  The adiabatic boundary losses are the sum of the CR energy flux over the outflow boundaries from cells with faces on those boundaries.  The diffusive boundary losses are the change in CR energy of boundary cells before and after the diffusion solver updates its solution.   }
\label{fig:CRterms}
\end{figure*}

Figure \ref{fig:CRterms} shows the evolution of the total CR energy in two representative simulations and the evolution of all the source and sink terms. These simulations create an outflow from the midplane. This is associated with a net expansion in the gravitationally stratified atmosphere and hence, a net adiabatic loss of CR energy.  It is this work that is the dominant loss.  In the model with the thermal runaway, the CR injection and adiabatic losses appear to be nearly in equilibrium, so CRs expand from the midplane almost as soon as they are injected.  Due to the injection of CR energy in a low density medium, it almost immediately expands adiabatically.  In models with peak driving, CR energy does not expand as readily and remains confined to the warm layer of gas for a longer period of time.  This also results in a different amount of CR energy being lost across the outflow boundaries, with it being higher in the thermal runaway models.

There are small differences in the amount of energy injected because there is a small difference in the SFR of these models, which sets this term.  The losses due to gas removal by star formation are cumulatively small, amounting to at most a few SN explosions worth of energy and subdominant in all cases.

The CR cooling is similar in both models. However, it is less than the adiabatic losses in both cases.  Thus, the CR energy contributes to the powering of outflows from the midplane.  Also in both cases, the hadronic cooling losses dominate over the Coulomb losses by a factor of more than 2.5.  While the thermal runaway model has more gas mass at high densities, which has a higher rate of CR-nucleon collisions, it occupies such a small volume that it only removes a small amount of CR energy.  In the peak driving run, the CR cooling approaches the adiabatic losses and arises from the warm neutral phase of gas that occupies a large volume in this run.

In summary, the reason the midplane thermal energy density dominates over the CR energy density in peak driving runs is the fragmentation of the midplane gas. This results in the dual effects of thermally heating the midplane gas to high temperatures and allowing for the more efficient outflow of CR energy deposited in the midplane, producing an overall drop in CR energy in the calculation.  In the peak-driving case, the stability of the warm gas layer results in overall lower midplane thermal temperatures and the less efficient escape of CR energy through $P$d$V$ work.  The only other sink term relevant in the midplane, CR cooling, is not efficient enough to alter the trend in the CR energy evolution, and the total amount of CRs in the calculation stabilises to a near equilibrium.  The overall result is a medium where CR energy density dominates over thermal energy density.

\subsection{Fast, slow, and no CRs: the impact of CR pressure and transport on the midplane structure}
\label{sec:kappa_diff}
\begin{figure*}
\centering
\includegraphics[width=\textwidth]{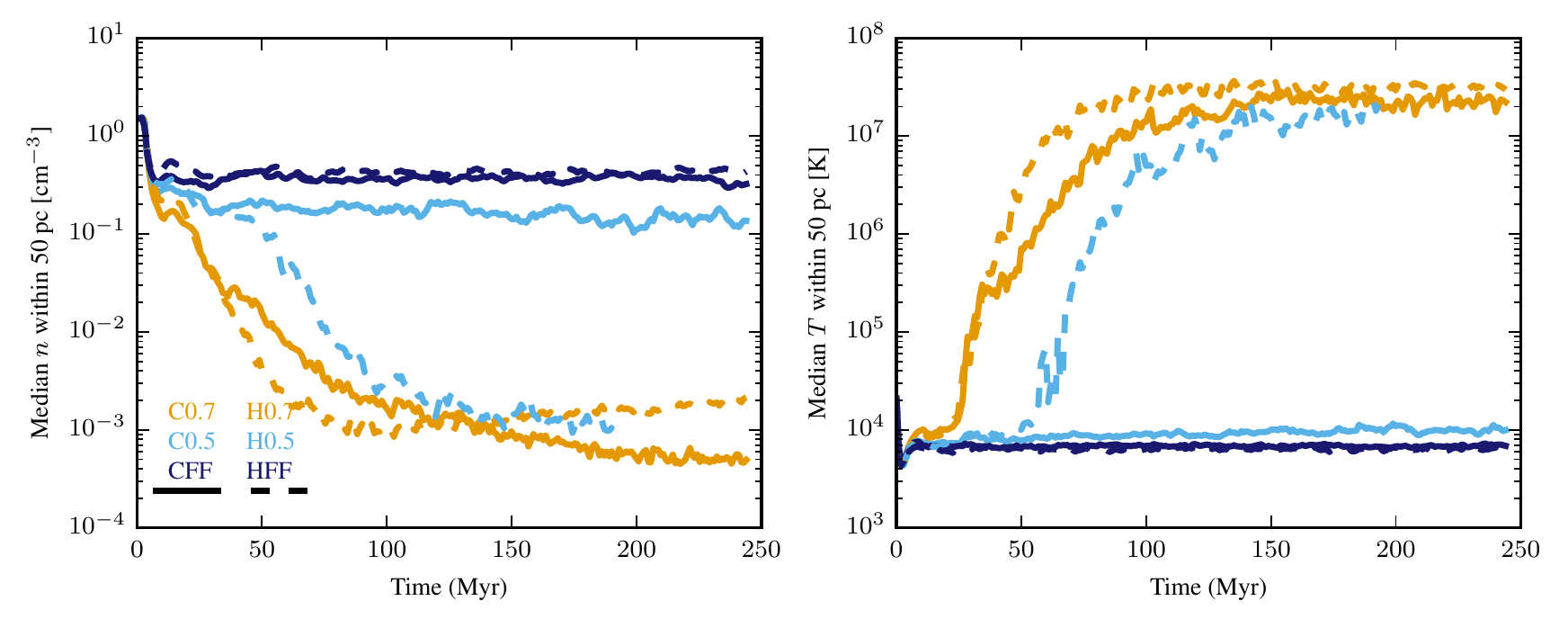}
\caption{Evolution of models with (solid) and without (dashed) CRs compared.  The median number density (left) and temperature (right) are shown as they are described in Figure \ref{fig:mix_evolution}.  The SN explosion placement models shown are the FF model (pure peak-driving) and two MIX models with \frand\ of 0.7 and 0.5.  The \frand$=0.5$ models evolve differently with and without CRs.  The presence of CRs is able to flip the model from the thermal runaway solution to the peak driving solution.}
\label{fig:rho_comp}
\end{figure*}

\begin{figure}
\centering
\includegraphics[width=\columnwidth]{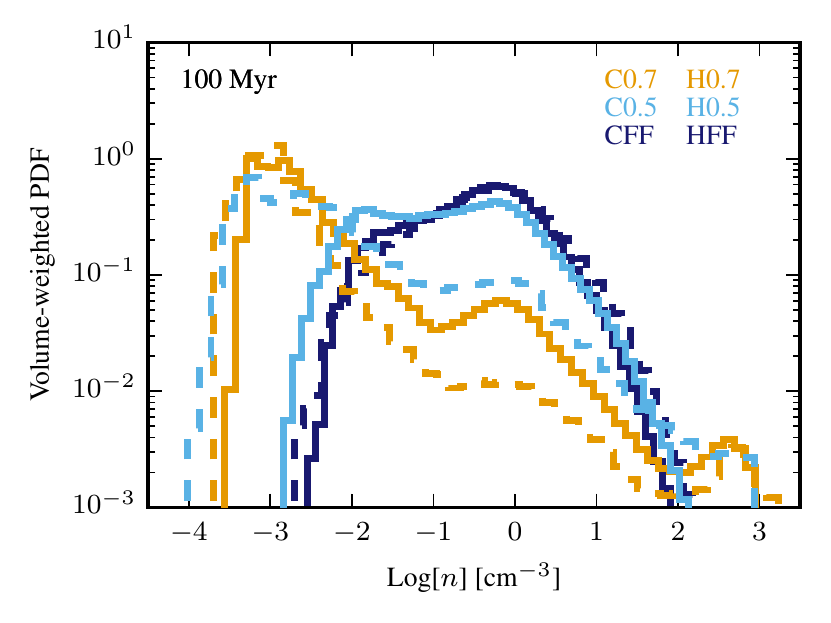}
\includegraphics[width=\columnwidth]{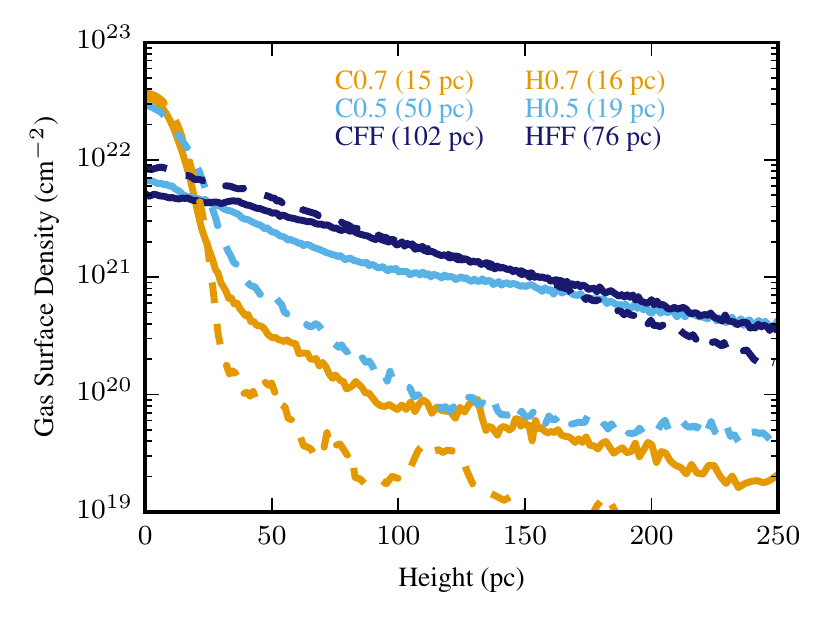}
\caption{Comparison of simulations with CRs (solid lines) and without CRs or MHD (dashed lines) for models with the same SN explosion placement.  The models shown are the same as in Figure \ref{fig:rho_comp}.  The same quantities are shown as in Fig. \ref{fig:profiles}: the distribution of gas densities within 50 pc of the midplane (top) and the vertical distribution of gas density (bottom).  The presence of CRs changes the behavour of models with \frand$=0.5$, but also thickens the amount of gas at heights about 100 pc in all models.}
\label{fig:profilecomp}
\end{figure}

The phase structure of the gas impacts the evolution of the CR energy as we have shown, but how do the CRs feed back on these structures?  How dynamically relevant are CRs to the phase structure evolution we have described?  

We have already considered the impact of removing CRs from the calculation altogether.  Figure \ref{fig:images} gives a visual impression of the impact of CRs in the calculation, contrasting projections of gas density in runs with and without CRs.  The models without CR diffusion do not track CR energy density and all SN energy is deposited in the form of thermal energy.  The only energies in the gas are therefore thermal and kinetic.  
Figures \ref{fig:rho_comp} and \ref{fig:profilecomp} show a comparison of mid-plane gas properties and evolution between these sets of runs.  
The FF models with and without CRs behave virtually identically, but the two selected MIX models behave differently.  As shown in Fig.~\ref{fig:rho_comp}, the model with 70 per cent random SNe evolves more quickly to a thermal runaway without CRs: the mid-plane density drops more quickly and the mid-plane temperature rises more quickly.  In the simulation with CRs, the median mid-plane density does drop below 0.01 cm$^{-3}$ and the temperature rises above $10^6$ K, it just takes somewhat longer to do so. By contrast, in the run with 50 per cent random SN explosions, the two cases evolve to different states.  The run without CRs undergoes a thermal runaway, but the run with CRs evolves to the peak driving state.  

Figures~\ref{fig:images}, \ref{fig:rho_comp}, and \ref{fig:profilecomp} show that the fraction of random SN events necessary to precipitate a thermal runaway is higher and occurs over a much narrower range of \frand\ when CRs are included.

\begin{figure*}
\centering
\includegraphics[width=\textwidth]{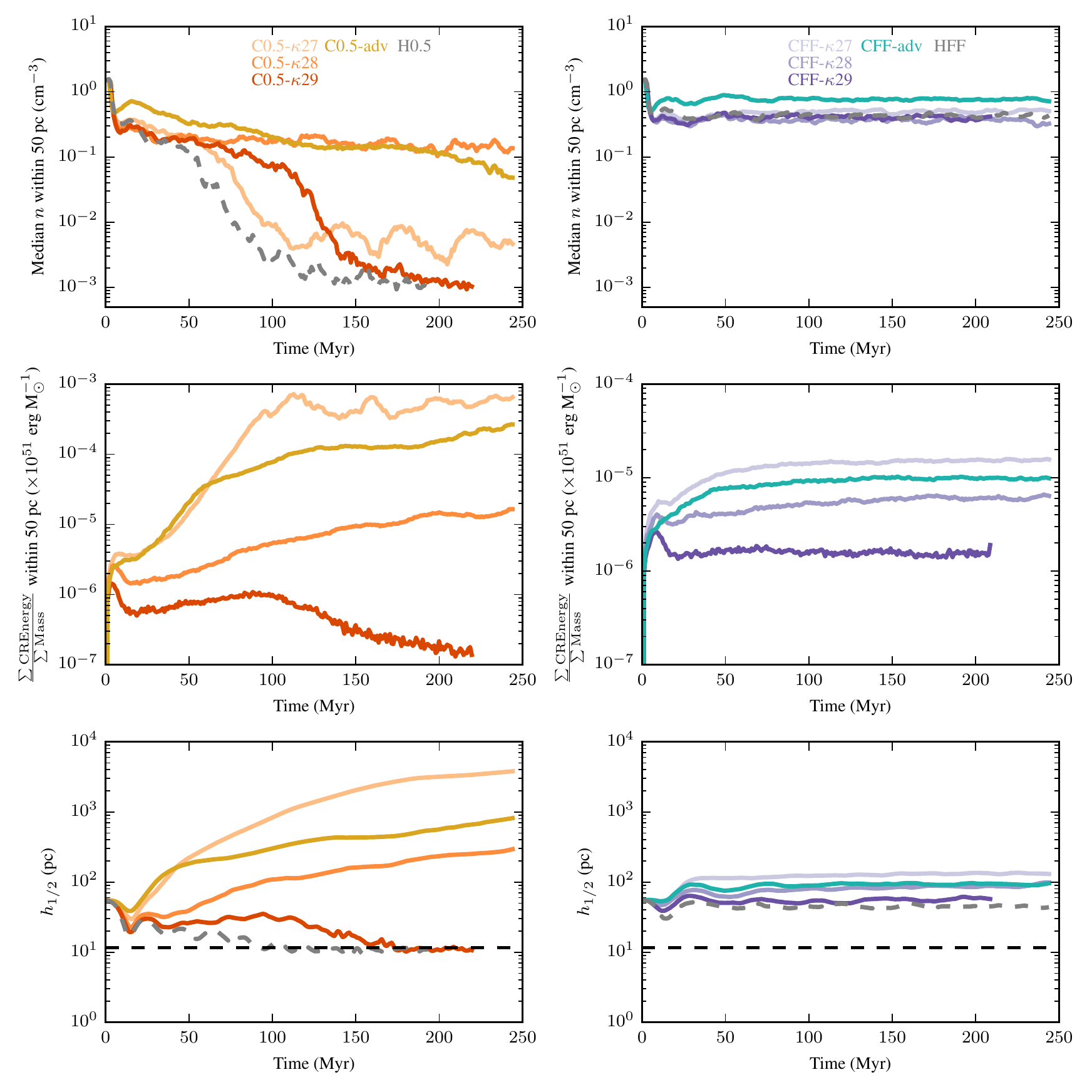}
\caption{Evolution of simulations run with varying $\kappa$, the global anisotropic diffusivity of the CR fluid.  
Top row: the median number density of the thermal gas within 50 pc of the midplane over time.  
Middle panel: the total CR energy within 50 pc of the midplane divided by the total gas mass in the same volume over time.
Bottom panel: the height containing half the original gas mass of the simulations.  
Two models for SNe placement are shown, the MIX0.5 model (left) and the FF model (right).  For both models the hydrodynamics-only model are shown with a dashed line.  Models with the CR fluid but evolved without diffusive transport are also shown.  As $\kappa$ decreases, the scale height of the midplane gas layer increases, but the phase structure of the midplane gas is still sensitive to the SN explosion placement model.
 }
\label{fig:kappa_evo}
\end{figure*}

In the setup tested here, the models with a 50 per cent random fraction represent a `tipping point' of the dynamical evolution and are therefore an interesting case to consider. This value of \frand\ represents a separatrix between two attractor solutions of the system: the thermal runaway and the self-regulated peak-driving solutions.  CRs change where this tipping-point occurs in addition to softening the vertical gas density gradient and causing the emergence of cooler and smoother galactic outflows in comparison to purely SN driven, ballistic outflows \citep{Girichidis2018,Simpson2016}.

The microphysics underlying the choice of the CR diffusion coefficient $\kappa$ is complex and depends on the specifics of CR plasma physics. A correct modelling of the plasma physics of CR transport would imply resolving the interplay of CR-driven instabilities \citep{Kulsrud1969,Shalaby2021} and various Alfv\'en wave damping processes \citep{Guo2008} so that the resulting level of Alfv\'en waves (in-)efficiently scatters CRs and thus set their transport speed (with a lower level of Alfv\'en waves implying less efficient CR scattering and thus faster diffusive transport). However, the large scale-separation between the CR gyroradius, where the wave-particle scattering happens, and the extent of galaxies precludes a direct plasma-dynamical modelling but instead argues for using a non-equilibrium fluid theory of CRs \citep{Zweibel2017,Jiang2018,Thomas2019}. 

In the self-confinement picture of CR transport, the CR diffusion coefficient $\kappa$ varies spatially and temporally as a result of the different balance of the Alfv\'en wave excitation and damping processes \citep{Thomas2019,Thomas2021a,Thomas2022a} and CRs are transported as a superposition of advection, streaming and diffusion as can be inferred from the radio brightness profiles of radio-emitting non-thermal filaments in the Galactic centre \citep{Thomas2020}. To study the sensitivity of our results to the diffusivity of the CR fluid, we have run the CFF and C0.5 models with different values of $\kappa$ that bracket our fiducial value, between $10^{27}$ and $10^{29}$ cm$^2$~s$^{-1}$. This enables us to assess whether the locus of the separatrix discussed above also depends on the value of the CR diffusion coefficient, and therefore how sensitive our ISM model would be to the details of the self-confinement picture. Our fiducial value of $\kappa$ is motivated by observations of the
fluxes of secondary-to-primary nuclei, such as the boron-to-carbon ratio measured at the Earth \citep[e.g.,][]{Evoli2020}.  We have also tested these models without diffusive transport ($\kappa = 0$), where the CR fluid is perfectly coupled to the thermal gas.

In our model, CRs diffuse along magnetic field lines and their diffusivity is set by the coefficient $\kappa$, which is the diffusion coefficient of the CR fluid in the direction of the local magnetic field according to Eq. \eqref{eq:cr_transport} (in our model, the physical diffusion coefficient is 0 in all directions perpendicular to $\bm{b}$ so any remaining perpendicular diffusion is of a purely numerical nature).  We have demonstrated how the presence of CR pressure in the midplane affects the fragmentation of the ISM.  Varying the diffusivity of the CR fluid by adjusting $\kappa$ will impact the energy density of CRs in the midplane, thereby influencing the ISM dynamics.

Figure \ref{fig:kappa_evo} shows the evolution of the midplane medium with different transport and SN explosion placement models.  
With the CFF model, runs with different $\kappa$ evolve to the same midplane density structure, and show little difference from the non-CR models.  
There is a modest change in the midplane height of gas, where $h_{1/2}$ increases with decreasing $\kappa$.  

With the C0.5 model, the midplane height of gas also increases with decreasing $\kappa$, but the effect is much more dramatic, over an order of magnitude.  The midplane density evolution shows that while the fiducial model asymptotes to the peak driving solution, the model with a larger $\kappa$ undergoes a thermal runaway, much like the model without CRs.  The model with a lower $\kappa$ undergoes a very different type of evolution.  The model with low $\kappa$ evacuates almost all of the gas from the midplane region because the CR pressure force acts for a longer time on the gas in comparison to the models with a larger  $\kappa$, creating a midplane dominated by low density gas, but without the dense molecular clouds found in runs that undergo a thermal runaway. 

\begin{figure*}
\centering
\includegraphics[width=\columnwidth]{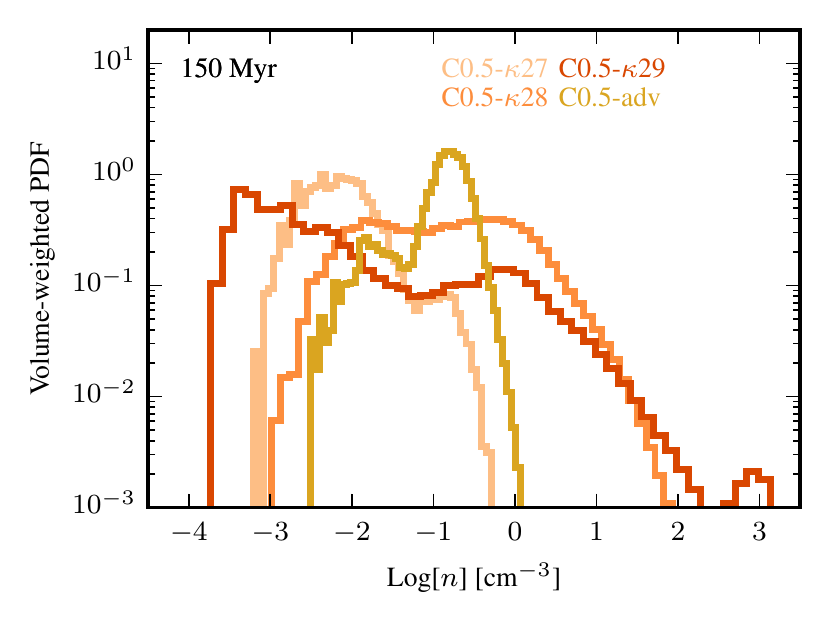}
\includegraphics[width=\columnwidth]{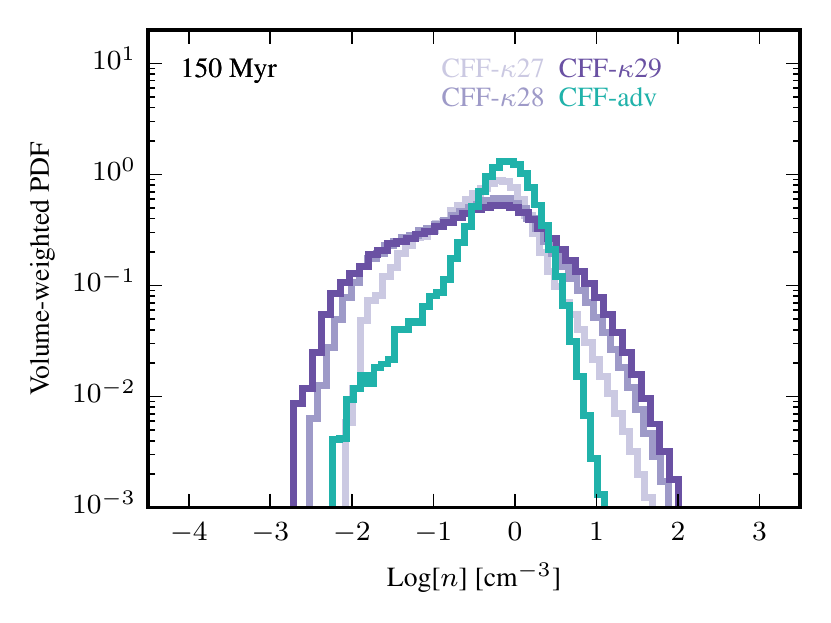}
\includegraphics[width=\columnwidth]{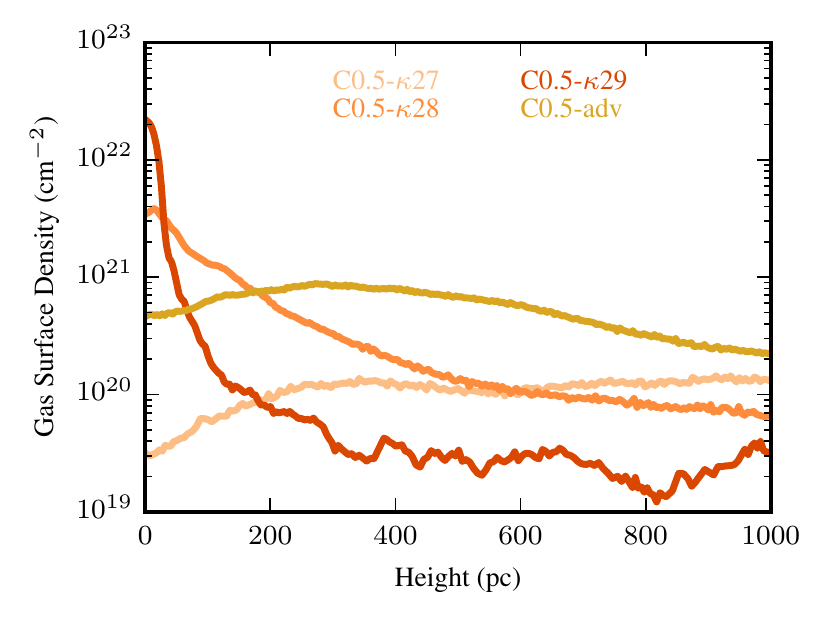}
\includegraphics[width=\columnwidth]{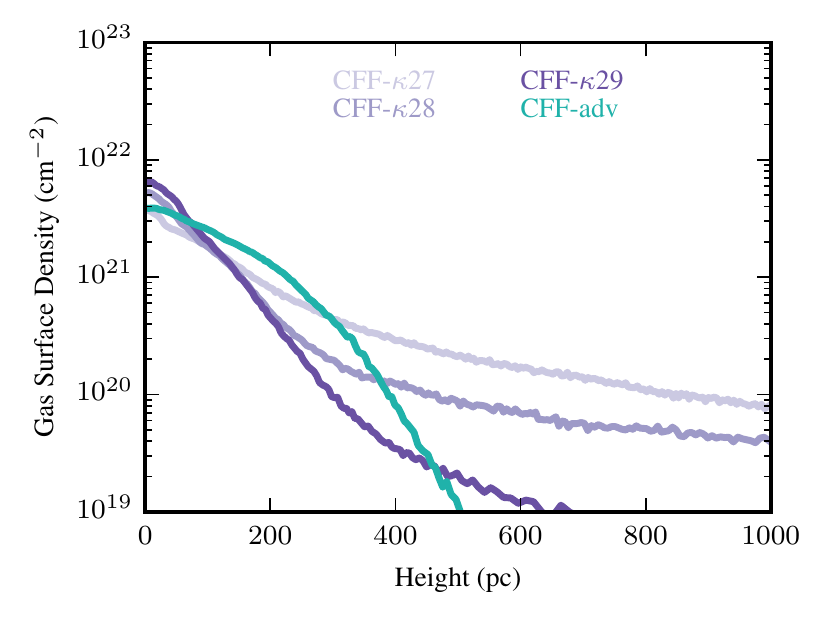}

\caption{Properties of models with different CR diffusivities after 150 Myr of evolution, including CR models with no diffusion.  Two SNe placement models are shown: one with SNe placed only in high density peaks (CFF, left) and one where 50 per cent of explosions are positioned randomly in the mid-plane (C0.5, right).  The fiducial CR model is run with $\kappa = 10^{28}$ cm$^{2}$s$^{-1}$.  Top: Volume-weighted probability density functions (PDFs) of the logarithm of the gas number density within 50 pc of the mid-plane.  Bottom: Surface number density of gas along the vertical direction.  The simulation volumes above and below the mid-plane are combined and the mid-plane is at height 0.  }
\label{fig:profileskappa}
\end{figure*}

This is clear in Fig.~\ref{fig:profileskappa} that shows the density and height distributions of midplane gas.  In the CFF model there is a smooth progression with $\kappa$ where the distribution of gas densities broadens with increasing $\kappa$, but the midplane layer thickens with more gas above 200 pc.  In the C0.5 runs, faster diffusion also results in a thermal runaway and less gas at larger scale heights (although, more than in the pure hydro model).  Slower diffusion prevents gas collapsing to densities higher than 1 cm$^{-3}$ and leads to an inverted vertical gas density profile below 200 pc.

CR models with no diffusive transport (where the CR energy density is perfectly coupled to the thermal gas) produce a thickened gas layer.  In C0.5-adv, the formation of dense clouds is suppressed as in the low diffusion case.  However, the density distribution is skewed to higher densities in the no diffusion case, resulting in a medium that is overall denser than the low diffusion case.  This is likely due to the lack of any diffusive transport, without which, gas remains confined to the midplane, albeit in a layer that is very puffy with a slight inversion in the central density profile.

From this analysis, we can conclude in the model tested here that the diffusivity of the CRs has more of an impact on the larger scale structure of the ISM (e.g. the scale height of the gas), and less on the midplane structure, unless the value of $\kappa$ is extremely high or low.  High values of $\kappa$ result in a model where CRs become unimportant in the midplane, and the evolution asymptotes to they model without CRs.  Low values of $\kappa$ can result in the suppression of dense clouds, except in the case where CR energy is continuously deposited in dense gas and CR cooling acts to remove CRs.

\subsection{Effective pressure support}
\label{sec:eff_pres}

\begin{figure*}
\centering
\includegraphics[width=\textwidth]{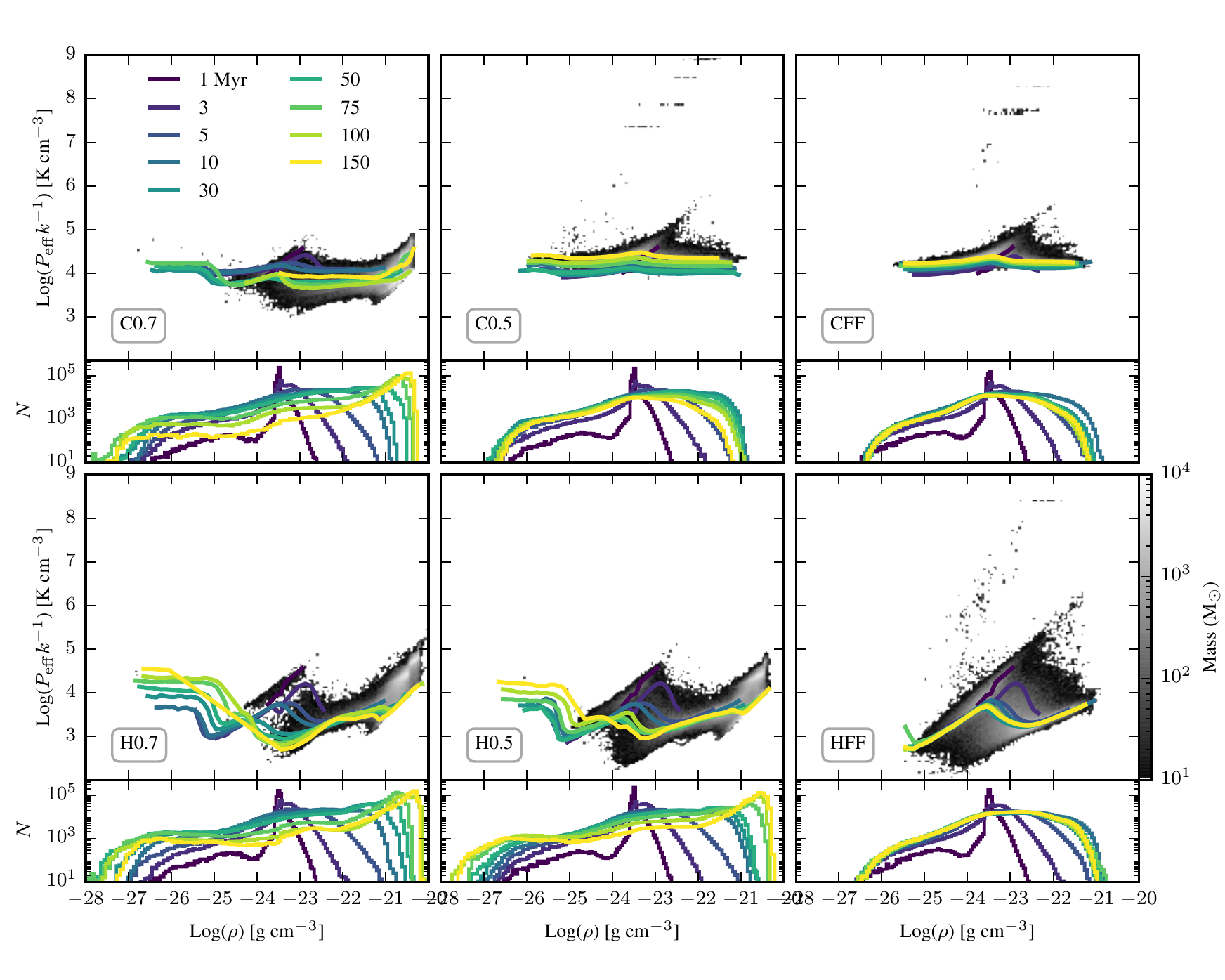}
\caption{The effective pressure evolution for three different SN explosion placement models run with (top row) and without (bottom row) CRs is shown.  Panels show the effective pressure vs.\ gas density after 100 Myr of evolution within 50 pc of the midplane in grayscale.  For the times indicated, curves following the median trend in this space are also shown.  Below the pressure panels, the distribution of cell gas densities are shown.}
\label{fig:Peff}
\end{figure*}

As we have seen, the inclusion of CRs can impede or slow the fragmentation of gas into different phases.  In models where the diffusive transport of CRs is slowed, the distribution of gas densities narrows.

To better understand this behaviour, we consider the effective pressure and how it influences the gas evolution.  The effective pressure is defined as
\begin{equation}
\label{eq:peff}
    P_{\mathrm{eff}} = P_{\mathrm{therm}} + P_{\mathrm{CR}}= (\gamma - 1) \varepsilon + (\gamma_{\mathrm{CR}} -1) \varepsilon_{\mathrm{CR}}.
\end{equation}
\noindent This definition of pressure combines thermal  pressure and CR pressure into a single measure of the pressure.  We neglect magnetic pressure because we found that it is almost always subdominant (see Fig.~\ref{fig:energyimages}).

Figure \ref{fig:Peff} shows the evolution of the distribution of effective pressure in our fiducial CR runs within 50 pc of the midplane (unlike Fig.~\ref{fig:phasespace} which shows the thermal pressure in a larger volume).  

This gross behaviour can be understood by considering the pressure support in the dense gas.  As gas cools, dense regions contract to maintain a constant thermal pressure (i.e. isobaric cooling).  In models with a sufficiently large reservoir of CRs, the CRs become important.  As dense gas contracts, CRs become adiabatically compressed which produces an outward CR pressure gradient force that if steep enough can stabilise the gas against collapse.  Therefore, in the presence of CRs, it is more difficult for gas to become denser from radiative cooling.

This can be seen in Fig. \ref{fig:Peff}.  The CRs provide a minimum effective pressure support in the gas that is constant across density (seen in C0.5 and CFF).  When gas radiatively cools in these models, pressure forces do not act to make gas more dense, because the CR pressure maintains an equilibrium.  

However, CRs cool and so need to be replenished to provide effective pressure support.  This occurs in the peak-driving case, where CRs are replenished by direct injection via SN explosions in dense gas.  In the thermal runaway case (e.g.\ C0.7 shown in Fig.~\ref{fig:Peff}), the only potentially efficient source of CR replenishment is diffusion of CRs from lower density gas into clouds.  But for our diffusivity, this process is not fast enough to overcome the radiative cooling.  The dense gas will therefore collapse further.  Ultimately, only our imposed Riemann pressure floor will halt collapse.

\section{Observational Signatures}
\label{sec:obs}
Here we explore predictions for the emission of $\gamma$ rays arising from interactions between the CR fluid and the thermal gas and the distribution of H I atoms that give rise to 21 cm emission.

\subsection{Hadronic $\gamma$-ray emission}
We follow \citet{Pfrommer2017b} in modeling $\gamma$-ray emission from interactions of CRs with the ISM.  
This model assumes $\gamma$ rays come from two sources: the primary channel from the decay of neutral pions created by inelastic collisions of CRs with thermal nucleons and the secondary channel from inverse Compton scattering of relativistic electrons.  
These processes are dependent on the energy spectrum of CRs, which we do not self-consistently simulate, but instead we assume a power-law CR momentum spectrum.  We focus on the primary production channel of $\gamma$ rays through pion decay, which is dominant for at the star formation rates in galaxies the mass of the Milky-Way and greater \citep{Werhahn2021b}.
The $\gamma$-ray emission from a cell (luminosity per unit volume) in an energy interval $[E_1,E_2]$ comes from integrating the source function $s_{\pi^0 - \gamma}(E_\gamma)$ of decaying neutral pions \citep[adopting the semi-analytical model of][]{PfrommerEnsslin2004}:
\begin{equation}
\Lambda_{\pi^0 - \gamma} =\int_{E_1}^{E_2} s_{\pi^0 - \gamma}(E) E \rmn{d}E
\label{eq:lambda}
\end{equation}
We compute the luminosity for each cell in the Fermi energy band (0.1-100 GeV).

We use an estimate for the $\gamma$-ray luminosity from \citet{Pfrommer2017b} who assume a power-law distribution for the momentum of CRs:
\begin{equation}
f(p) = C_\rmn{p} p^{-\alpha} \theta(p-q),
\label{CRdist}
\end{equation}
\noindent where $p =  P_\rmn{p}/m_\rmn{p} c$ is a dimensionless proton momentum ($m_\rmn{p}$ is the proton mass and $c$ is the speed of light).
The function $\theta$ is the Heavyside step function and $q$ is the low-threshold momentum limit.  We adopt $q=0.5$ to account for the fact that low-energy CRs are efficiently cooling via Coulomb and ionization interactions \citep{Werhahn2021a,Werhahn2021b}. 
We choose a fixed power-law slope of $\alpha = 2.05$.  A more natural choice for the power-law slope would seem to be $\alpha = 2.2$ \citep{Haggerty2020,Caprioli2020}, corresponding also to the slope that maximises pion decay emission in the Fermi band.  However, this slope, as applied in this analytic model, has been shown to be effectively degenerate with the choice of CR SN injection efficiency \citep{Pfrommer2017b}.  Our chosen injection efficiency of 10 per cent, while within acceptable limits, has recently been shown to likely be an overestimate, as SN acceleration models that account for the magnetic field-shock front obliquity favour a fraction closer to 5 per cent when integrated over the entire remnant, and independent of the magnetic coherence length \citep{Caprioli2014,Pais2018}.  To account for this, we have chosen the value of $\alpha$ appropriate for this factor of two higher injection efficiency, i.e. one that results in a factor of two lower pion decay emission.

The normalisation factor $C_\rmn{p}$ is determined for each cell by equating the simulated energy density of CRs to the integrated energy density from our assumed $f(p)$ (which is a good approximation for the CR spectrum above the kinematic threshold $p=0.78$ of the hadronic reaction; see \citet{Girichidis2022} for a simulated CR spectrum in a galaxy simulation):
\begin{equation}
\varepsilon_{\rmn{CR}} = \int^\infty_0 f(p) E(p) \rmn{d}p
\end{equation}
\noindent where $E(p) = (\sqrt{1+p^2} -1) m_\rmn{p} c^2$ is the kinetic proton energy.  \citet{Werhahn2021b} demonstrate (in their Appendix~B) that this analytical model reproduces the result of a more detailed numerical integration that includes the energy-dependent hadronic cross section with an accuracy of 10 per cent. 

Under these assumptions, the source function and hence the luminosity density of $\gamma$ rays are proportional to the energy density of CRs times the number density of target nucleons for the hadronic reaction: $\Lambda_{\pi^0 - \gamma} \propto \varepsilon_\mathrm{CR} n_{\rmn{n}}$.  The spatial distribution of gamma rays is therefore highly dependent on the density structure of the thermal gas (because $n_\rmn{n}=\rho/m_\rmn{p}$) and the speed with which CRs are transported from their birth sites.

\begin{figure*}
\centering
\includegraphics[width=\textwidth]{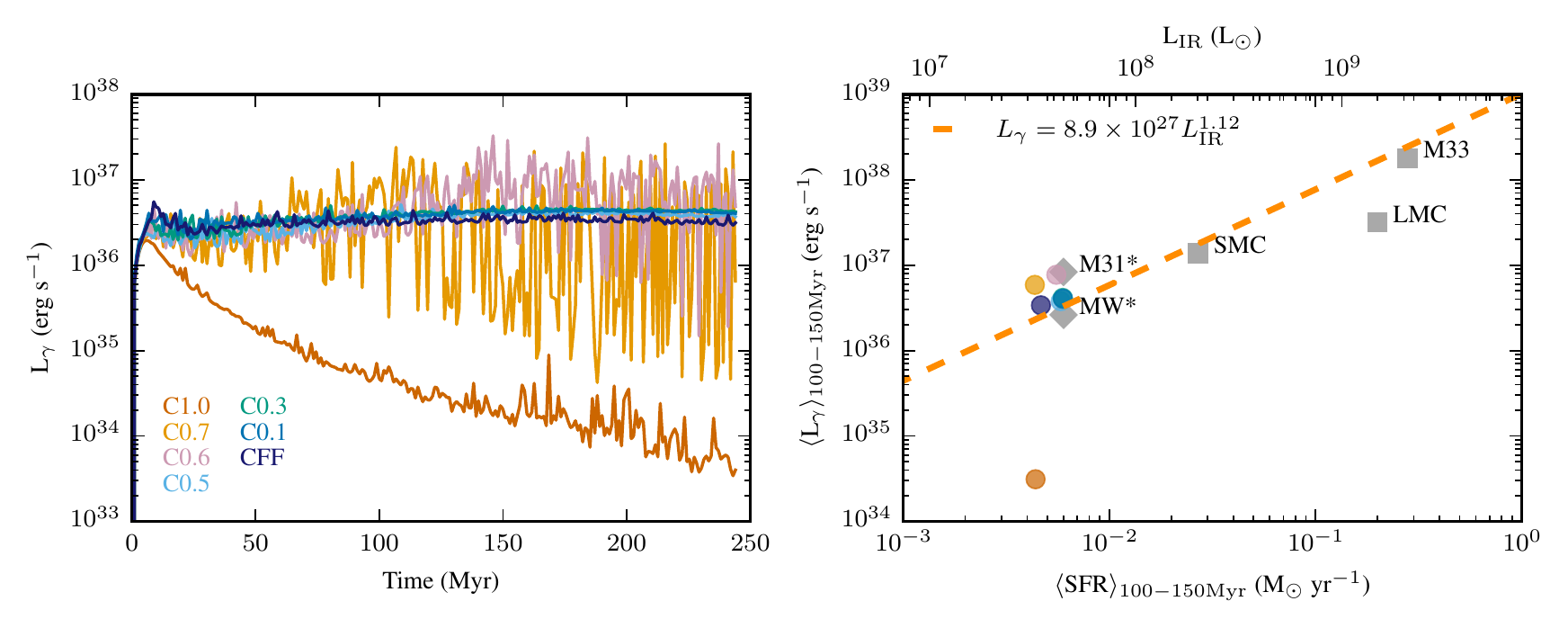}
\caption{Gamma-ray emission arising from pion decay in the Fermi band of 0.1 to 100 GeV.  Left: Time evolution of the total gamma ray luminosity of each simulation volume.  Thermal-runaway models demonstrate a higher degree of variability than peak driving runs. Right: The average $\gamma$-ray luminosity averaged between 100 and 150 Myr for all the runs.  An extrapolation of the best fit line to galaxies observed both with Fermi and in the infrared is shown (dashed orange line) assuming a relation between infrared emission and SFR from Kennicutt.  Observed values for the SMC, LMC, and M33 are shown (squares) and values for the MW and M31 are shown scaled to a SFR of $6 \times 10^{-3}$ \Msun yr$^{-1}$ (diamonds), corresponding to the local patch of ISM simulated here.  }
\label{fig:gamma_evo}
\end{figure*}

The total $\gamma$-ray luminosity in the Fermi band from pion decay in the patch of ISM simulated here can be computed by summing the emission from all cells in the simulation volume.
The evolution of this luminosity is shown in Fig.~\ref{fig:gamma_evo}, which shows that the time variability of $\gamma$-ray emission differs between the two solutions discussed in Sec.~\ref{sec:mainresults}. 
In the thermal runaway models, the emission is highly variable.  
It is especially variable in models with a non-zero number of SNe in high density peaks, like C0.7 and C0.6.  
The models that asymptote to the peak driving solution have very minimal time variability.
The overall level of emission is very similar between all models except for the purely random placement model, C1.0.  That model has an overall luminosity an order of magnitude lower than the other models.

The reasons for these differences is revealed in Fig.~\ref{fig:energyimages}.  
This figure shows that fast diffusion in the midplane creates a roughly spatially uniform distribution of CR energy throughout the midplane.
The level and variability of $\gamma$-ray emission is therefore set by the distribution and volume-filling factor of nucleons.  
In peak-driving models where the midplane volume is mostly filled by warm neutral gas and where there are smaller gas density perturbations, the emission from recent SNe is not very different from the background $\gamma$-ray emission coming from this warm gas.

In the thermal-runaway models, most of the midplane volume is filled by very low density ionized gas, and most nucleons are found in the molecular clouds that occupy a very small volume of the midplane. The background $\gamma$-ray emission is therefore small compared to the emission that arises when SNe occur near dense clouds.  
The variability of the thermal-runaway models are therefore a result the concentration of target nucleons in molecular clouds and the MIX SNe placement model.
Periods of high emission arise when a recently injected SNe coincides spatially with a molecular cloud and low emission when randomly injected events dominate.  
Freshly injected CR energy near a molecular cloud does not remain spatially concentrated for long since we have no model for variable diffusion \citep[such as][]{Semenov2021}.  The energy quickly diffuses, causing the emission of $\gamma$ rays to drop in the region of the molecular cloud and therefore the entire box.
In C1.0 where all SNe are injected randomly, the coincidence of a recent SNe and a molecular cloud is very rare (by construction) so C1.0 has less variability and an overall lower level of emission.
The runs that have the highest degree of variability are runs where there is a large \frand, but with some non-zero seeding of SNe in dense gas where the placement of SNe in molecular clouds light them up in $\gamma$ rays.

The relation between $\gamma$-ray luminosity and SFR is also seen in Fig.~\ref{fig:gamma_evo}.  Galactic-scale $\gamma$-ray emission is observed to correlate with FIR emission over several orders of magnitude for systems where Fermi has been able to make measurements, from the SMC to Arp 220 \citep{Ackermann2012,Rojas-Bravo2016}.  Our models do not probe a range of star-forming environments so we can only probe this relation at a single SFR; however, we see that the average level of $\gamma$-ray emission does lie on the observed galactic-scale trend.  The one exception is C1.0, which as we have explained has a lower level of $\gamma$-ray emission.  

\begin{figure}
\centering
\includegraphics[width=\columnwidth]{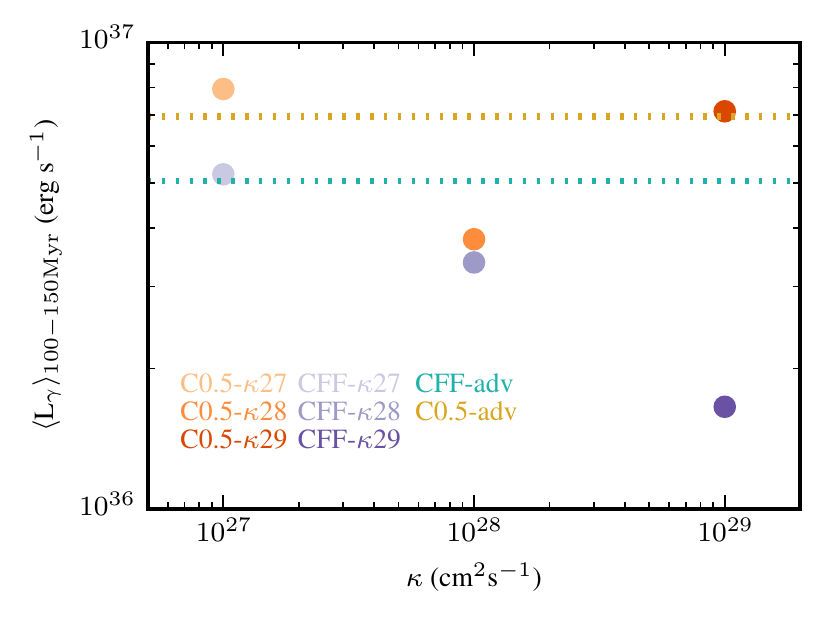}
\caption{Gamma-ray emission arising from pion decay in the Fermi band of 0.1 to 100 GeV averaged between 100 and 150 Myrs.  Models run with different CR diffusivities $\kappa$ are shown with both the CFF model (purple points) and the C0.5 model (orange points).  Models run with CRs but no diffusive transport (advection only models) are shown as dotted lines for the CFF model and C0.5 model. }
\label{fig:kappa_gamma}
\end{figure}

Figure \ref{fig:kappa_gamma} shows how the $\gamma$-ray luminosity varies with the CR diffusion coefficient, $\kappa$.  The luminosity can vary by a factor of several with $\kappa$, with lower $\kappa$ models trending toward higher luminosities, especially for the CFF models, thus matching the pure CR advection models with $\kappa=0$.  This is due to the accumulation of CR energy in the midplane that results from the slower diffusion.  
It is also the case, however, that the presence of CRs impacts the amount of dense gas in the midplane, the gas that is most luminous in $\gamma$ rays.  In C0.5-$\kappa$29, the fast diffusion of CRs reduces the amount of CRs relative to mass in the midplane (see Fig.~\ref{fig:kappa_evo}), however, the reduced CR pressure allows more dense gas to form (see Fig. \ref{fig:profileskappa}).  Even with the reduced CR energy density in the midplane, the enhanced dense gas means that the model has on average an enhanced $\gamma$-ray luminosity.

\subsection{Atomic Hydrogen}

\begin{figure*}
\centering
\includegraphics[width=\textwidth]{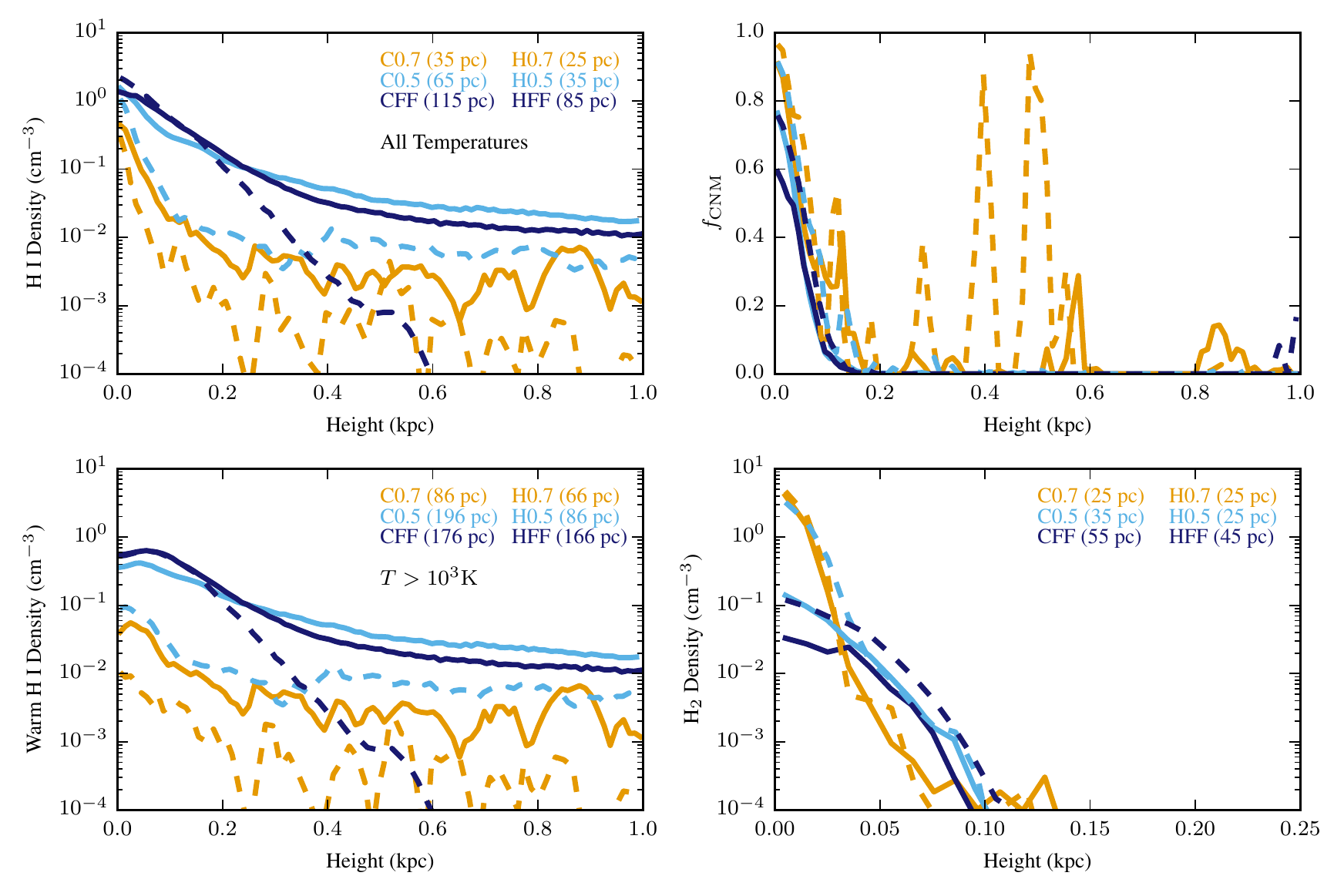}
\caption{The vertical distributions of different species of neutral hydrogen shown after 100 Myr of evolution for simulations with CRs (solid lines) and without CRs (dashed lines).  Top left: The total number density of H I atoms vs. height.  The e-folding scale heights of the density are indicated.  All gas regardless of temperature is included in this plot.  Top right: The fraction of H I atoms in cold gas ($T < 10^3$~K), representing the mass fraction of the CNM.  Bottom left: The number density of H I atoms in gas that is warm ($T > 10^3$~K).  This gas corresponds to the optically thin WNM.  Bottom right: The number density of H$_2$ molecules.}
\label{fig:hydrogen_profs}
\end{figure*}

\begin{figure}
\centering
\includegraphics[width=\columnwidth]{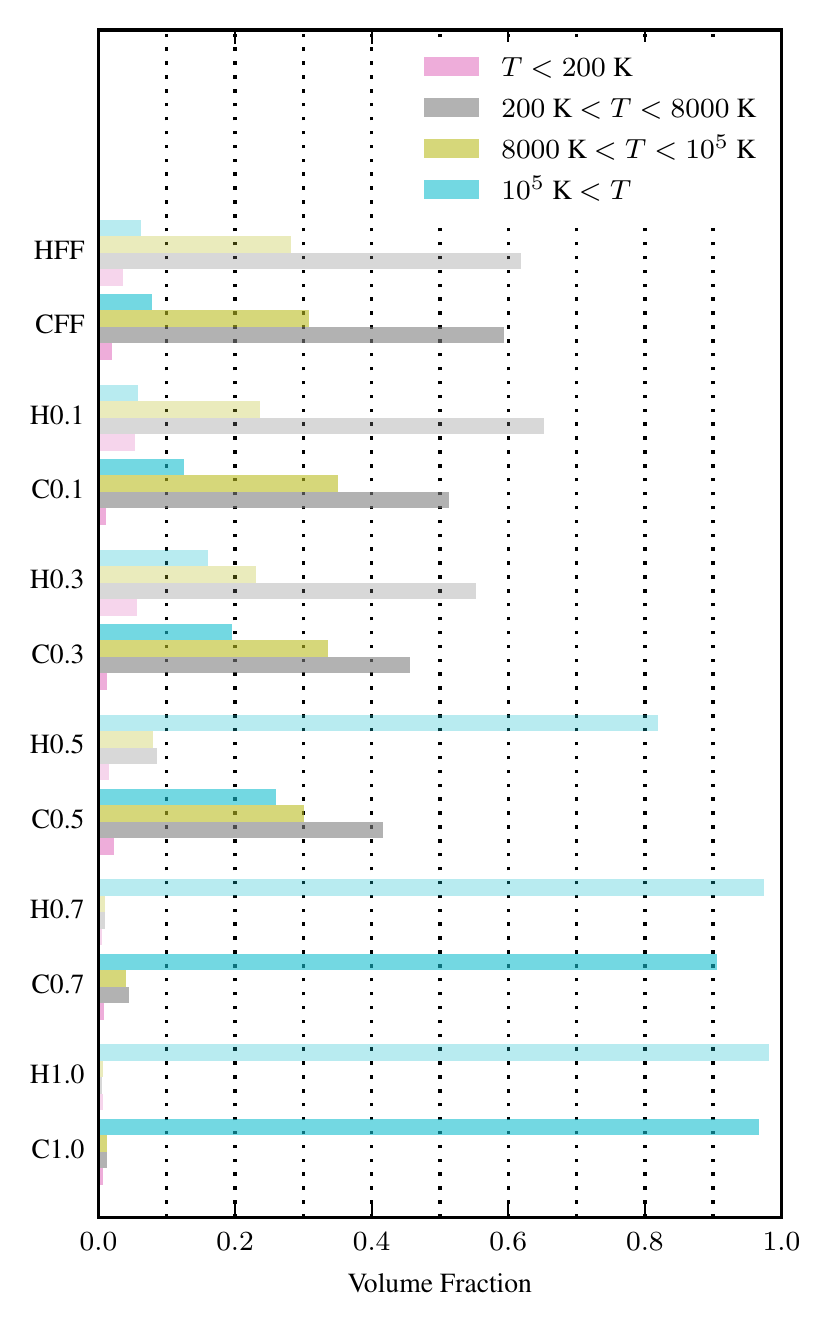}
\includegraphics[width=\columnwidth]{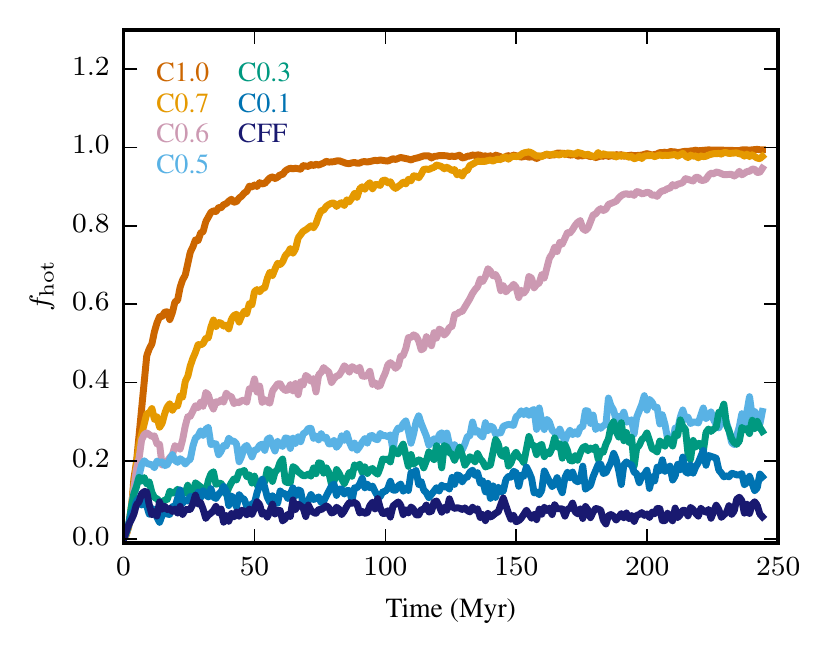}
\caption{Top: Fraction of the midplane volume (within 50 pc of the midplane) occupied by gas in different temperature ranges.  Simulations with (dark colors) and without (light colors) CRs are compared with varying SN explosion placement models.  These volume fractions are averaged between 95 and 104 Myr. Bottom: The evolution of the volume fraction of gas with temperatures greater than $10^5$ K within 50 pc of the midplane.  }
\label{fig:volratio}
\end{figure}

Figure~\ref{fig:hydrogen_profs} shows the vertical column density of H I and H$_2$ gas in models with and without CRs.  The scale height, temperature, and composition of neutral gas varies between injection models, but also changes with and without CRs.

Some of these differences are due primarily to the differing solutions to which these models evolve, but some differences seem to be due to the presence of CRs irrespective of the midplane's ISM structure.  Models that evolve to the peak-driving solution have more vertically extended H I and lower surface densities of H$_2$ when CRs are included.  Thermal runaway models have less H I gas above the midplane and very concentrated H$_2$ layers.  The peak surface density of H$_2$ in all the thermal runaway models saturates at a value of $\sim 3\times 10^{22}$ cm$^{-2}$.  This constant value is likely due to the Riemann pressure floor imposed in dense gas.

Part of the extended H I gas in the CR runs may be part of the base of a CR driven outflow, which we will explore in future work.  The overall thickness of the H I layer varies between approximately 20 pc (C1.0) and 150 pc (C0.1) in these runs while it is typically less than 50 pc in the no CR runs.  We define these heights as the 1 e-folding height, i.e. where the gas surface density drops by 1/e.  

In all models, the central midplane H I gas is dominated by cold gas with temperatures less than $10^3$ K.  The fraction of H I that is composed of this cold neutral medium (CNM) varies from 60 to nearly 100 per cent as shown in Fig. \ref{fig:hydrogen_profs}.  Even at a height of 100 pc, the CNM fraction ($f_\rmn{CNM}$) can vary between 20 and 40 per cent.  These high fractions indicate that a substantial amount of H I gas may be optically thick in the midplane, which would impact the column density of H I emission at 21 cm.

The height of the MW's 21 cm emission at the solar circle is approximately 150 pc as defined as the half width half maximum \citep{KalberlaDedes2008}, larger than the upper range of our models.
However, given the large central CNM fractions in our simulations, even up to a height of 100 pc, the scale height of the H I volume density should be regarded as a lower limit for any inferred height of the 21 cm emission because a large portion of the cold H I will be optically thick.  We can consider the warm H I, with temperatures greater than $10^3$ K, as providing an upper limit to the scale height of the 21 cm emission, as this warm H I will certainly be optically thin.  Figure~\ref{fig:hydrogen_profs} shows that the e-folding heights of the warm H I atoms are substantially larger than the total H I.  The height of the gas layer of this optically thin H I is typically greater than 150 pc for peak-driving runs.  For thermal runaway models, the WNM is also more extended, but does not approach 150 pc in scale height.

We also present in Fig. \ref{fig:hydrogen_profs} the distribution of H$_2$.  It should be noted, however, in order to capture a converged H$_2$ component, much higher resolution is needed \citep{Joshi2019}.  Zoom simulations of molecular clouds suggest that at our resolution, the underestimate of the molecular gas fraction is of order a factor of two \citep{Seifried2017}.

Our models have demonstrated that CRs enhance the thickness of the H I layer.  With the set of physics we consider in this study, CRs appear to thicken the H I layer to a degree that brings models, with varying SN placement assumptions, in better agreement with the MW's 21 cm emission.

\section{Discussion}
\label{sec:discussion}

The ISM system examined here includes many of the physical effects found in the real ISM of the MW, but there are also many effects we have not considered in this study.  Comparing the solutions we have identified here in this model to the state of the MW's ISM can give us some insights into this other physics.  

Figure \ref{fig:volratio} shows the fraction of the midplane volume occupied by gas of different temperatures.  The transition between the peak driving and thermal runaway solutions is evident in these fractions.  In the MW, measurements of gas in these temperature ranges are dependent on different observables that are sensitive to density and ionization state, in addition to temperature.  Generally, the hot ionized medium (HIM) is thought to fill approximately 50 per cent of the volume.  The warm ionized medium (WIM), gas of roughtly $10^4$~K found in H II regions, comprises about 20-30 per cent of the volume, and the warm neutral medium (WNM), gas of several thousand K and traced by warm H I gas, comprises about 30-40 per cent.  Cold gas, with temperatures $\le 100$~K and traced by cold H I (the CNM) and molecules, is of order one per cent of the volume \citep[e.g.][]{Cox2005,Draine}.

In our model, the HIM is either too volume-filling (occupying over 80 per cent of the volume) or not volume-filling enough (less that 30 per cent of the volume).  Other studies have found that a set of clustered SN explosions or SN explosions from dynamically formed sink particles are able to match the volume fraction of the HIM \citep[e.g.][]{Rathjen2021}.  We find that as the models in the peak driving phase get closer to tipping over to a thermal runaway, the volume fraction occupied by the HIM can pass through 50 per cent, but it is not a stable state with the physical model tested here.

The WNM and WIM are completely suppressed in simulations with a thermal runaway, unlike the MW where combined, they comprise about half the ISM volume.  In the peak driving models, they occupy an appreciable amount of the volume as seen in Fig. \ref{fig:volratio}.  We separate the gas into temperature ranges here only, but we see that the cooler gas component (grey bands with $200 \mathrm{\ K} < T < 8000 \mathrm{\ K}$) is the dominant component of the medium, typically 40 per cent of the volume when \frand$= 0.5$ and trending to 60 per cent in the FF models.  The component at higher temperatures ($8000~\mathrm{K} < T < 10^5~\mathrm{K}$) seems to be more consistent in its volume fraction across peak driving runs ($\sim25$ per cent).  This gas appears to be cooling SN remnants placed in the dense gas.  They are constantly replenished by the injection model as they constantly cool away due thermal radiative cooling.  

This origin for the WIM is different from the MW where it is typically associated with H II regions produced by massive stars.  While the simulations presented here have an interstellar radiation field attenuated by dense gas, they do not include a dynamic radiation field that includes radiation transport.  \citet{Kannan2020} in models similar to those presented here tested the effects of photoionzation and radiation pressure with full radiation transport and found an enhancement in the supply of warm gas ($10^3$ K $< T < 10^5$ K).  The combination of radiation and CR transport will likely combine in a non-linear fashion which we will explore in future work.  However, the simulations of \citet{Rathjen2021} that include both effects suggest that the inclusion of radiation is especially important for capturing the warm phase.

Beyond radiation transport, the models here neglect by construction many effects of disc dynamics and the cosmological environment of the ISM.  Turbulence plays an important role in the development of dense gas and in our model is seeded primarily by SN explosions, however, disc models suggest that gravity driven turbulence may be more consistent with observations of gas velocity dispersions \citep{Krumholz&Burkhart}.  Disc shearing likely plays an important role in the creation of dense gas, even at the resolution considered here \citep[e.g.][]{Smith2014}.  The stratified gas layer considered here allows for the outflow of gas from the midplane, but there is no CGM included that would be a potential reservoir of gas inflow to the disc or a mediator for fountain flows that arise from the disc but fall back \citep{Tumlinson2017}. 

While these models reach a high resolution of several pc in the midplane (see Fig. \ref{fig:resolution}) and this is likely sufficient to capture the cooling of SN energy, we do not have the resolution to capture the development of SN remnant bubbles self consistently and in the development of the top layers of the midplane, the steep gradient in gas density produces a steep gradient in cell sizes (although limited by our adjacent volume limit refinement criterion).  Numerical mixing in these top layers may result in a suppression of dense structures or different turbulent properties.

We have presented a description of the range of behaviours for the ISM in the presence of a diffusive CR fluid with a constant $\kappa$.  Models of SN remnants suggest that non-resonant streaming instabilities \citep[e.g.][]{Bell2004} may lead to a decrease in CR diffusion away from SN remnants \citep{Schroer2022}. \citet{Semenov2021} test on much larger scales the impact of spatially varying CR diffusion in galactic discs and find that suppressing the diffusion of CRs away from birth sites enhances disc stability and prevents the fragmentation of gas. In addition, the interplay of resonant CR streaming instabilities \citep{Kulsrud1969,Shalaby2021} and various wave damping processes should establish an effective CR propagation speed that varies spatially and temporally. \citet{Thomas2022b} found that while the CR diffusion coefficient reaches a steady state in most galactic environments, CR transport itself appears not to reach a steady state and therefore, cannot be well described by either CR streaming, CR diffusion or a combination of both paradigms but instead obeys non-equilibrium transport mediated by resonant Alfv\'en waves.

These plasma effects may shift the transition \frand\ in our model between the peak driving and thermal runaway solutions.  This is because confining CRs injected in dense gas within clouds longer would require a less frequent replenishment of fresh CRs to alter the cloud's collapse.  However, if $\kappa$ is allowed to increase away from the sites of recent SN explosions, CRs may escape the medium so quickly that effects like the enhanced H I layer thickness may not materialise.  We note that while we do not vary $\kappa$, the CR diffusion time does vary with gas density as shown in Fig. \ref{fig:timescales}.  It is actually fastest in high density gas, which is likely the opposite trend that would result from a smaller $\kappa$ near sites of SN explosions coincident with dense gas.

\section{Conclusions}

The models we have described have allowed us to test the behaviour of a diffusive CR fluid under specific astrophysical conditions, i.e.\ the application of SN explosions to a magnetised ISM at a prescribed rate in following a prescribed spatial distribution.  We have varied these astrophysical variables to test the tolerance of the collisional and diffusive fluids that comprise our simulated ISM system.  We have neglected many important effects for the evolution of the ISM: clustered SN explosions, disc dynamics, spatially and time varying radiation transport, and ISM-CGM interactions.  This was intentionally adopted to provide a simpler system with which to focus on the unadulterated impact of CRs at high resolution where we are capturing resolved SN explosions.  This approach lays a foundation on which we can better interpret a richer set of astrophysical effects in future work.

Our main results are as follows:
\begin{itemize}
    \item We have systematically varied the amount of SN energy injected into high-density gas vs.\ random locations, and consistent with previous works, models with large amounts of SN energy injected into dense gas evolve to a `peak-driving' solution where warm gas is volume filling.  Models with more SN energy injected into random locations develop a `thermal runaway' where hot gas becomes volume filling.
    \item The transition between these two solutions is mediated by CRs.  As seen visually in Fig. \ref{fig:images}, when CRs are present, the medium only fragments when more SN energy is coupled to random locations (i.e.\ into low density gas).  We see in Figs.~\ref{fig:rho_comp} and \ref{fig:profilecomp} that CRs can entirely prevent the development of the thermal runaway over the 200 Myr run time of our simulations with some placement models.
    \item We examine the partition of energy within the midplane of our simulations and find on global scales, turbulent and thermal energies are typically greater than CR energy when the medium is dominated by $T > 10^6$ K gas and less when it is dominated by warm gas ($T < 10^4$ K) as shown in Fig.~\ref{fig:edensities}.  However there can be a a great deal of spatial variation on pc scales in the distribution of thermal, kinetic, and magnetic energies while fast diffusion of the CR fluid means it can maintain an approximately spatially uniform distribution within the midplane (see Fig.~\ref{fig:energyimages})
    \item The evolution of CR energy within simulations is a balance between injection through the SN model and a variety of sink terms, the dominant one being adiabatic losses from the expanding midplane gas (see Fig.~\ref{fig:CRterms}).  Hadronic and Couloumb losses can be significant as well especially in simulations dominated by warm gas.
    \item We tested different values of $\kappa$, the diffusivity of the CR fluid, in the presence of two different SN placement models.  The comparison of these simulations is shown in Figs.~\ref{fig:kappa_evo} and \ref{fig:profileskappa}.  We find that fast diffusion results in models similar to the hydrodynamics only models.  Slow diffusion results in the build up of CR energy in the midplane.  The result of this build up can vary depending on the SN placement model, with some models creating pressure gradients so strong that gas is evacuated from the midplane.
    \item We examined the evolution of the effective pressure in the midplane and show that CRs impose a minimum pressure floor in the medium across densities (Fig.~\ref{fig:Peff}).  In cases where evolution of the thermal pressure is significant, this pressure floor can break, allowing for the fragmentation of the medium due to SN explosions.  Otherwise, the uniformity in $P_{\rm eff}$ across density, prevents gas from changing phase and fragmenting significantly.
    \item We make predictions for the $\gamma$-ray emission from the hadronic pion decay.  We find that the time averaged emission in most of our SN placement models with the fiducial diffusivity of $10^{28}$ cm$^2$ s$^{-1}$ matches the MW emission (see Fig.~\ref{fig:gamma_evo}).  Variability in the $\gamma$-ray emission is driven by the proximity of SN explosions to dense gas, which we demonstrate with our models that explicitly vary this.
    \item We show that the presence of CRs enhances the volume density of neutral atomic hydrogen at heights several 100 pc above the midplane in all SN placement models, as shown in Fig. \ref{fig:hydrogen_profs}.  The scale heights of H I volume density is consistent with the scale heights of 21 cm emission when considering most of the midplane H I simulated here will be optically thick.
    \item We examined the volumetric partition of gas of different temperatures and found that in all the SN placement models we have tested, it is difficult to sustain a volume fraction of hot gas of 50 per cent.  This may be to physical effects not considered here such as radiation transport or variable CR diffusion or streaming.
\end{itemize}

The stratified box model tested here is a flexible platform on which to conduct parameter space studies and test physical ideas in a controlled fashion as we have done here.  We have clearly demonstrated that CR transport and pressure are important effects that should be included in studies of the ISM going forward.  We plan to extend this work to also explore how CRs combine with radiation transport in the future, along with varying CR transport (including streaming and diffusion), and effects from runaway stars.

\section*{Acknowledgements}
This work was completed in part with resources provided by the University of Chicago’s Research Computing Center and the Enrico Fermi Institute.  We also acknowledge support from the Klaus Tschira Foundation.  This research was supported in part by the National Science Foundation under Grant No. NSF PHY-1748958. CP acknowledges support by the European Research Council under ERC-CoG grant CRAGSMAN-646955.

\bibliographystyle{mnras}
\bibliography{references}

\appendix
\section{Pressure Limiter}
\label{appendix}

Here we describe how the Riemann solver pressure limiter impacts the behaviour of the model.  
This model introduces a variable parameter $\rho_{\rm{max}}$ which is the maximum density of a Jeans stable cloud (see Eq.~\ref{eq:rhomax}). In gas above this density, the gravitational self-collapse of gas is prevented by a limit placed on the pressure used in the Riemann solver, thereby limiting the fluxes between cells.  Without this limiter, gas will collapse to arbitrarily high density because of gas self-gravity and radiative cooling.  

The collapse of gas to a small volume can be prevented by halting cell refinement at a minimum cell volume, a common approach in adaptive mesh refinement (AMR) codes and one employed in \citet{Simpson2016}.  However, we found that the extreme gradients resulting from over-massive cells produced spurious amplification of the simulated magnetic field, and made this an unworkable solution for our problem.

Other similar simulations have side-stepped this issue by not including gas self gravity \citep[e.g.][]{Hill2012}, and indeed this limits the collapse of clouds, but may impact the fragmentation of the medium on certain scales \citep[e.g.][]{Walch2015}. Another approach is to provide a sink for collapsing gas in the form of star particles that will shift gas mass to a collsionless component that will be subject to gravitational softening.  However, with the resolution we employ here, only a few per cent of collapsing gas should go into stars, making this an inappropriate solution to halt collapse.  Ultimately, neglecting low-temperature cooling below $10^4$ K would also be effective in preventing the collapse of dense gas \citep[e.g.][]{Creasey2013}, but of course, this does not allow for the study of the structure of the ISM on sub-kiloparsec scales as is our goal.

The main goal the simulations presented here is to understand the impact of CRs on the scale of the local solar neighborhood, a region of roughly a kpc in scale, and not to study the evolution of molecular clouds, for which we do not include necessary physical effects or appropriate resolution.

\begin{figure}
\centering
\includegraphics[width=\columnwidth]{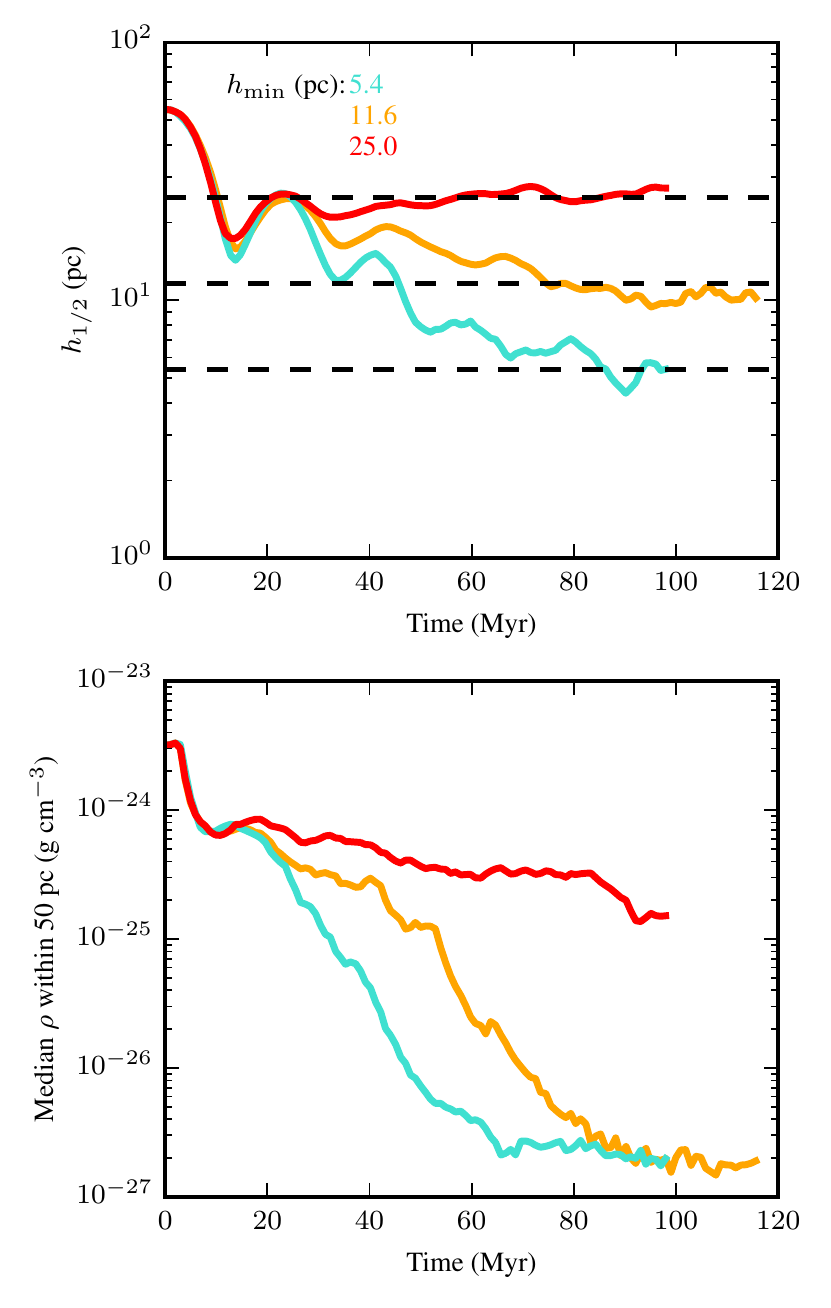}
\caption{Evolution of three boxes simulated with only hydrodynamics (no CRs and no magnetic fields) where 50 percent of SN explosions occur in random locations and 50 per cent take place in high-density gas (model H0.5 in the text).  These models all have different values of \rhomax: the fiducial value used in the main text of \rhomax$\sim 2 \times 10^{-22}$ (gold), a \rhomax\ ten times the fiducial value (cyan) and one tenth of the fiducial \rhomax\ (red).  The top panel shows the height containing half the original mass of the box and the bottom panel shows the volume-weighted median gas density within 50 pc of the midplane.  The top panel includes dashed lines showing the radii of a Jeans stable cloud with average densities at these different values of \rhomax.}
\label{fig:rhomax_evo}
\end{figure}

\begin{figure}
\centering
\includegraphics[width=\columnwidth]{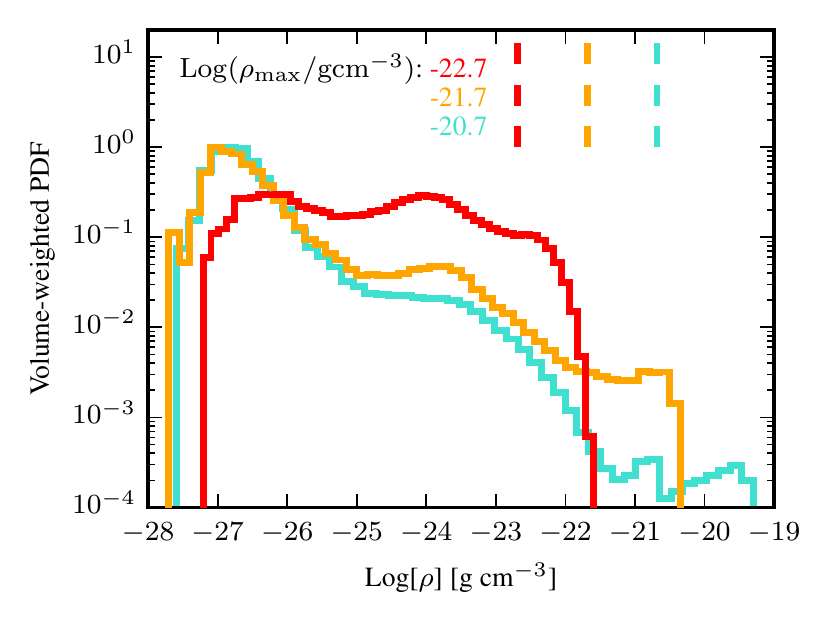}
\caption{The volume-weighted distribution of gas densities after 100 Myr of evolution for the three models shown in Fig. \ref{fig:rhomax_evo}.  The value of \rhomax \ for each model is indicated.}
\label{fig:rhomax_distr}
\end{figure}

\begin{figure*}
\centering
\includegraphics[width=\textwidth]{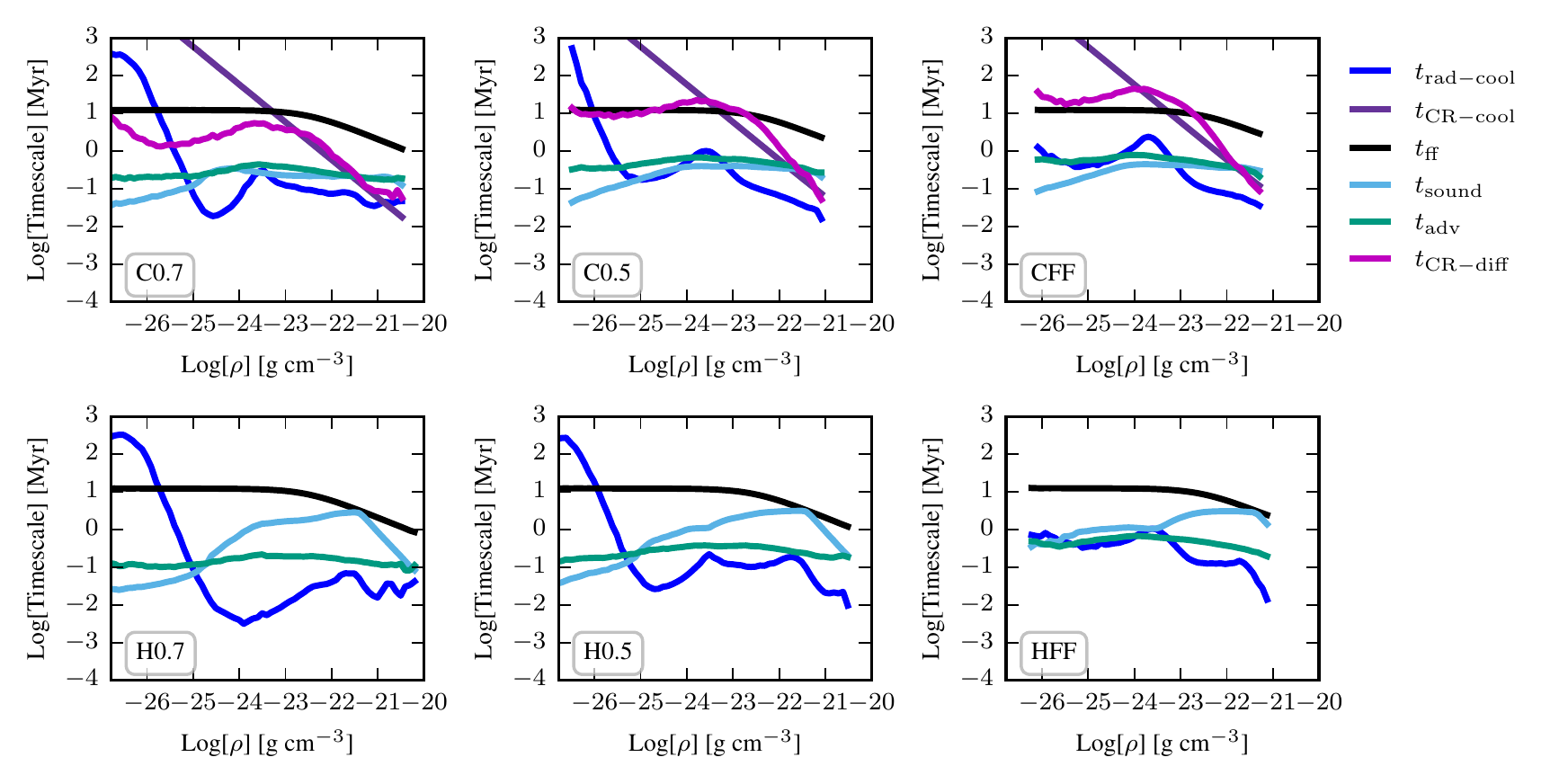}
\caption{Timescales of the ISM within 50 pc of the midplane after 100 Myr of evolution in models with varying SN explosion placement (left to right) and with and without CRs (top and bottom).  The median timescale is plotted as a function of gas density.  The timescales presented are the effective cooling time that accounts for CRs and their hadronic cooling ($t_{\rm{eff-cool}}$), the gravitational free fall time ($t_{\rm{ff}}$), the sound crossing time ($t_{\rm{sound}}$), the turbulent crossing time ($t_{\rm{turb}}$), the CR diffusion time ($t_{\rm CR-diff}$), and the CR cooling time ($t_{CR-cool}$). The models C0.7, H0.7, and H0.5 have undergone a thermal runaway and their midplane volume is dominated by hot, low density gas.  The models C0.5, CFF, and HFF remain in the peak driving solution and their midplane volume is dominated by warm gas.  The transition between the warm and cold phase of the ISM occurs around $10^{23}$ g cm$^{-3}$ and the timescales around this density are the most relevant to the problem.}
\label{fig:timescales}
\end{figure*}

Figure \ref{fig:rhomax_evo} shows the behaviour of test runs with a higher and lower values of \rhomax\ at fixed mass resolution ($m_\rmn{t} =10$ \Msun).  In effect, what varies in these models is the radius of a Jeans stable cloud in the simulation, where runs with low values of \rhomax\ are only able to capture larger clouds.  Figure \ref{fig:rhomax_evo} shows that indeed, gas in these runs collapses to a layer with a height equivalent to the size of a Jeans stable cloud of this mass.

This model is intended to halt the collapse of dense clouds in the absence of a full model for star-formation and pre-SN feedback.  
However, if applied in extremity, this pressure limiter can act as a form of subgrid feedback that impacts gas at densities below \rhomax.
Figure \ref{fig:rhomax_distr} shows this effect.  The model with the lowest \rhomax\ tested has more gas at densities of $10^{-24}$ g cm$^{-3}$ because the low value of \rhomax\ prevents gas at these densities from collapsing to higher densities.  Our fiducial value of \rhomax\ allows for this collapse and results in a gas density distribution resembling that of a model with a higher \rhomax\ but lacking the the tail of densities above $10^{-20}$ g cm$^{-3}$.

These models' large scale behavior largely depends on whether gas can transition from the WIM around a temperature of $10^4$ down to temperatures around 100 K of the cold neutral medium.  Figure~\ref{fig:phasespace} shows this transition happens around a density of $10^{-23}$ g cm$^{-3}$.  A choice of \rhomax\ too close to this value will impede this transition and affect the development of the WNM and CNM.
This in turn will impact the porosity of the medium and the amount of gas available to be driven from the mid-plane in outflows.  

We have presented tests with a feedback model where 50 per cent of SN explosions occur in random locations and 50 per cent in dense gas.  With our fiducial model parameters, this model undergoes a thermal runaway.  This model without CRs is the model closest to tipping over to the peak driving solution.  It behaves similarly to the model with a value of \rhomax\ 10 times larger.  They reach a similar average log-density and have similar gas density distributions up to densities of \rhomax.  With a value of \rhomax\ 0.1 times smaller, the medium does not undergo the same type of collapse.  However, the distribution of gas is different from the peak driving models seen in Fig. \ref{fig:profilecomp}.  Gas is more extended in height by a factor of several and it's distribution in density is broader, but more uniform up to \rhomax.  

The choice of \rhomax, therefore does have the ability to significantly alter the model's behaviour, but here we have demonstrated our fiducial value of \rhomax\ for our chosen mass resolution allows this transitional case to still undergo a thermal runaway.  Of course, in models with a large number of SN explosions occurring in dense gas, the injected SN energy make the pressure limiter irrelevant because the injected energy is effective in countering cooling and prevents gas from reaching \rhomax.  Models that have a high \frand\ are most sensitive to the details of the pressure limiter.

\section{Timescales}
\label{appendix_timescales}

Useful quantities to consider in the evolution of the ISM  are the timescales of various gas processes shown in Fig. \ref{fig:timescales}.  The timescales we consider are the gravitational free fall time ($t_{\rm{ff}}$, see Equation \ref{eq:ff}), the radiative cooling time of the thermal gas ($t_{\rm{rad-cool}}$),  the sound crossing time ($t_{\rm{sound}}$), the advective crossing time ($t_{\rm{adv}}$), the CR cooling time ($t_{\rm{CR-cool}}$), the CR diffusion time ($t_{\rm{CR-diff}}$)and the SNe heating time ($t_{\rm{SN-heat}}$).

The radiative cooling time is an output of the chemistry and cooling network used here \citep{Glover&MacLow2007a,Glover&MacLow2007b,Smith2014} that includes atomic and molecular hydrogen cooling along with metal line cooling for a metallicity of 1 \Zsun.  

The CR cooling time is defined as the time it would take to reduce the CR energy in a cell by an e-folding as a result of hadronic and Coulomb losses:
\begin{equation}
    t_{\rm{CR-cool}} =  \frac{\varepsilon_{\rm{CR}} \rho}{\Lambda_{\rm{hadr}} + \Lambda_{\rm{Coul}}},
\end{equation}
\noindent where $\Lambda_{\rm{Coul}}$ and $\Lambda_{\rm{hadr}}$ are the volumetric loss rates of CR energy from these processes (assuming a steady-state CR distribution).

The CR diffusion time is related to the CR pressure gradient in the direction of the local magnetic field ($\mathbf{b}$) and the diffusivity $\kappa$:
\begin{equation}
    t_{\mathrm{CR-diff}} = \left(\frac{P_\mathrm{CR}}{\mathbf{b}\bcdot\bnabla P_\mathrm{CR} }\right)^2 \kappa^{-1}.
\end{equation}

We also consider the sound crossing time and the advective crossing time, which are the ratios of the cell's approximate diameter $\Delta x$ derived from its volume $V_i$ and the sound speed of the medium in the cell ($c_{i,\rm{s}}$) or the bulk velocity of the cell ($v_i$), respectively:
\begin{equation}
t_{\rm{sound}} = \Delta x c_{i,\rm{s}}^{-1}
\end{equation}
and
\begin{equation}
t_{\rm{adv}} = \Delta x v_i^{-1}
\end{equation}
\noindent where $\Delta x = (6 V_i/\pi)^{1/3}$.  The sound speed depends on the pressure within the cell:
\begin{equation}
c_{i,\rm{s}} = \left(\frac{\frac{5}{3} P_{i,\rm{therm}} + \frac{4}{3} P_{i,\rm{CR}} + B_{i}^2/(8 \pi)}{\rho_i} \right)^{1/2},
\end{equation}
\noindent where $P_{i,\rm{therm}}$ is the thermal pressure and $P_{i,\rm{CR}}$ is the CR pressure.  Models that do not include magnetic fields or CRs only include the thermal pressure term.  In the case of non-zero-magnetic fields (that we evolve with the ideal MHD assumption), we adopt the maximum speed of the fast magnetosonic wave that is driven by magnetic pressure.
Models with CRs tend to have a higher total pressures due to the contribution of the CR fluid.  This results in higher sound speeds and lower sound crossing times.

Figure \ref{fig:timescales} compares these timescales as a function of density in runs with different SN placement models and with and without CRs.  The density where gas transitions from the warm phase to the cold phase is around $10^{23}$ g cm$^{-3}$ as shown in Fig. \ref{fig:phasespace}.

In all models at the relevant densities, the median free-fall time is typically too long to be relevant, but that does not mean that on small scales gas self-gravity cannot play a role in the collapse of gas.  What is perhaps more interesting is the role of gas motions, which can be a source of dynamical heating, but can also cause the collapse of gas through shock compression or shock collisions.  The sound crossing time becomes shorter with the addition of CRs due to their impact on the sound speed in the gas (it becomes faster).  The cell-crossing (or advection) time remains unchanged, indicating that the typical gas bulk motions remain unchanged with the addition of CRs.  

The CR diffusion time becomes shorter in denser gas.  In models with a thermal runaway and very dense gas, it becomes shorter than the radiative cooling time.  In this situation, the CR diffusion removes CR energy from dense gas faster than the gas has a chance to cool, indicating CRs leave before they can impact the thermal gas evolution.

\label{lastpage}

\end{document}